\documentclass[twocolumn,onecolappendix]{aastex63}

\usepackage{makecell}
\usepackage{xcolor}
\usepackage{comment}
\definecolor{dgreen}{RGB}{26,148,49}
\definecolor{xlinkcolor}{cmyk}{1,1,0,0}
\hypersetup{linkcolor=xlinkcolor,citecolor=xlinkcolor,urlcolor=xlinkcolor}

\usepackage{amsmath,amsthm,amssymb,amsfonts}
\usepackage{graphics}
\usepackage{listings}
\usepackage{enumitem}
\usepackage{url}
\usepackage{appendix}
\usepackage{hyperref}
\usepackage{xcolor}
\usepackage{siunitx}
\usepackage{lineno}

\usepackage{mathtools}

\usepackage[super]{nth}
\usepackage{outlines}
\usepackage{enumitem}

\newcommand{\N}{\mathbb{N}}
\newcommand{\Z}{\mathbb{Z}}

\begin{document}
\title{The SPHEREx Satellite Mission}


\author[0000-0002-5710-5212]{James~J.~Bock}%
\affiliation{Department of Physics, California Institute of Technology, 1200 E. California Boulevard, Pasadena, CA 91125, USA}%
\affiliation{Jet Propulsion Laboratory, California Institute of Technology, 4800 Oak Grove Drive, Pasadena, CA 91109, USA}%
\email{jjb@astro.caltech.edu}%

\author{Asad M. Aboobaker}
\affiliation{Jet Propulsion Laboratory, California Institute of Technology, 4800 Oak Grove Drive, Pasadena, CA 91109, USA}%

\author[0000-0001-9082-9354]{Joseph~Adamo}%
\affiliation{Department of Astronomy and Steward Observatory, University of Arizona, 933 North Cherry Ave, Tucson, Arizona 85721, USA}%

\author[0000-0001-9674-1564]{Rachel~Akeson}%
\affiliation{IPAC, California Insitute of Technology, MC 100-22, 1200 E California Blvd Pasadena, CA 91125, USA}%

\author{John M. Alred}
\affiliation{Jet Propulsion Laboratory, California Institute of Technology, 4800 Oak Grove Drive, Pasadena, CA 91109, USA}%

\author{Farah Alibay}
\affiliation{Jet Propulsion Laboratory, California Institute of Technology, 4800 Oak Grove Drive, Pasadena, CA 91109, USA}%

\author[0000-0002-3993-0745]{Matthew~L.~N.~Ashby}%
\affiliation{Center for Astrophysics $|$ Harvard \& Smithsonian, Optical and Infrared Astronomy Division, Cambridge, MA 01238, USA}%

\author[0000-0002-2618-1124]{Yoonsoo~P.~Bach}%
\affiliation{Korea Astronomy and Space Science Institute (KASI), 776 Daedeok-daero, Yuseong-gu, Daejeon 34055, Republic of Korea}%

\author[0000-0001-7665-5079]{Lindsey~E.~Bleem}%
\affiliation{High-Energy Physics Division, Argonne National Laboratory, 9700 South Cass Avenue., Lemont, IL, 60439, USA}%

\author{Douglas Bolton}
\affiliation{Jet Propulsion Laboratory, California Institute of Technology, 4800 Oak Grove Drive, Pasadena, CA 91109, USA}%

% ...
%

\begin{abstract}
SPHEREx, a NASA Explorer satellite launched on 11 March 2025, is carrying out the first all-sky near-infrared
spectral survey.  The satellite observes in 102 spectral bands from 0.75 to 5.0 $\mu m$ with a resolving
power ranging from $\lambda/\Delta\lambda = 35 - 130$ in 6\farcs2 pixels. The observatory obtains
a $5 \sigma$ depth of 19.5 - 19.9 AB mag for $0.75 < \lambda < 3.8 ~\mu m$ with $\lambda/\Delta\lambda \sim 40$ and 17.8 - 18.8 AB mag for $3.8 < \lambda < 5.0 ~\mu m$ with $\lambda/\Delta\lambda \sim 120$ after mapping the full sky four times over two years.  Scientifically,
SPHEREx will produce a large galaxy redshift survey over the full sky to constrain the amplitude
of inflationary non-Gaussianity.  The observations will produce two deep spectral maps near the ecliptic
poles that use intensity mapping to probe the evolution of galaxies over cosmic history.  By mapping
the depth of infrared absorption features over the Galactic plane, SPHEREx will comprehensively
survey the abundance and composition of water and other biogenic ice species in the interstellar medium.
The project will release initial data rapidly in the form of spectral images, and specialized data products
over the life of the mission as the surveys proceed.  The science team will also produce
spectral catalogs of planet-bearing and low-mass stars, solar system objects, and galaxy clusters 3 years after
launch.  We describe the design of the instrument and spacecraft, which flow from
the core science requirements.  Finally, we present an initial evaluation of the satellite’s in-flight
performance and key characteristics.

\end{abstract}



\section{introduction}
\label{sec:introduction}

\begin{figure}
\centering
    \includegraphics[width=\linewidth]{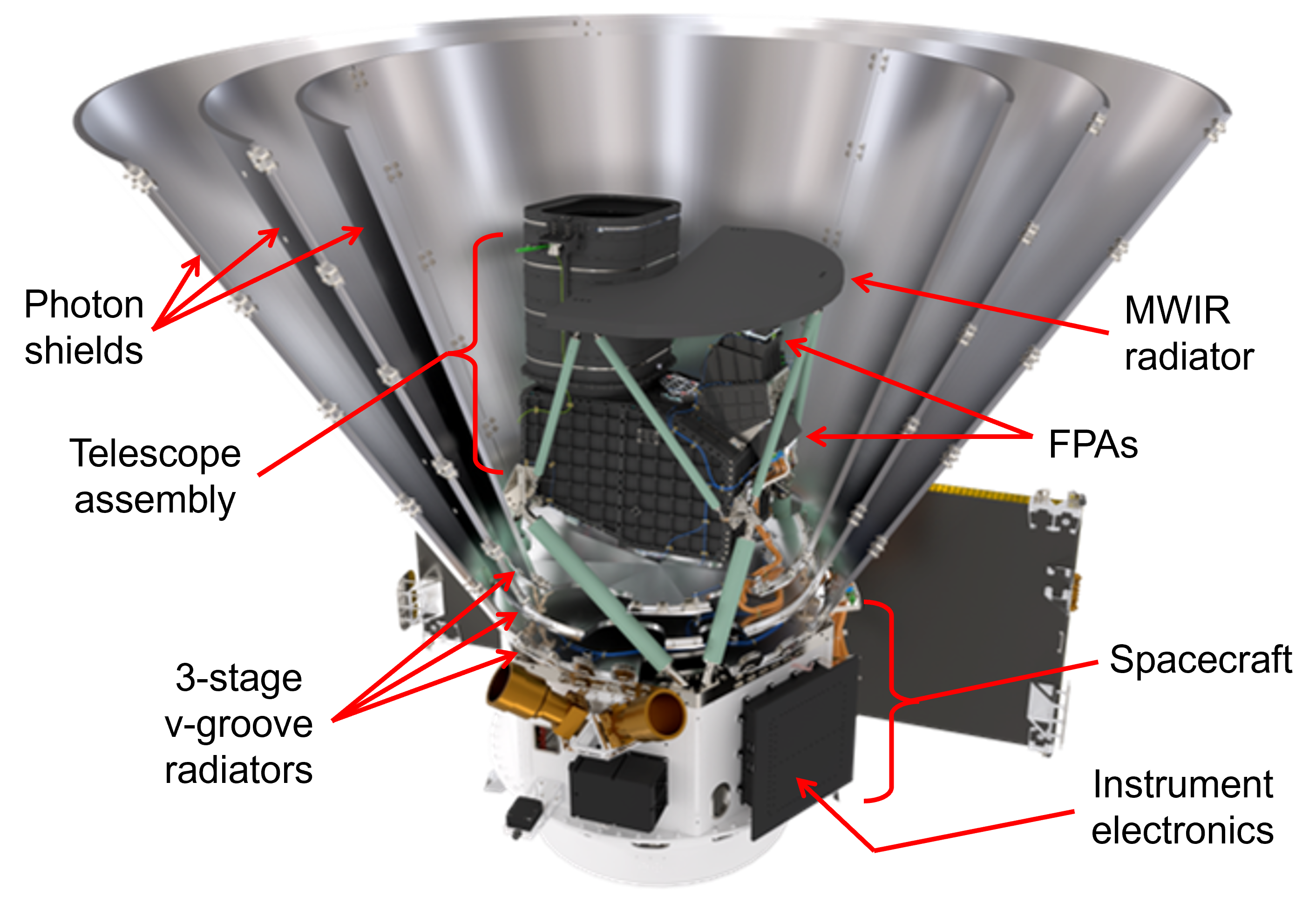}
    \caption{Diagram of the SPHEREx observatory, showing (from top to bottom) the three conical
    photon shields with a cutaway view, the telescope, radiator panel for the mid-wave infrared (MWIR)
    detectors, the MWIR and short-wave infrared (SWIR) focal plane arrays (FPAs), and the 3-stage V-groove radiator assembly
    with penetrating bipods for supporting the telescope assembly.  The BAE spacecraft at the bottom provides
    pointing, power, and telemetry and houses the instrument readout electronics.}
    \label{fig:spherex}
\end{figure}

The Spectro-Photometer for the History of the universe, Epoch of Reionization, and Ices Explorer
(SPHEREx) is designed to map the entire sky at near-infrared wavelengths
($\lambda = 0.75 - 5 \mu m$) in low-resolution spectroscopy
($\lambda / \Delta \lambda \sim 35 - 130$) with moderate 6\arcsec ~angular
resolution.  SPHEREx (see Figure \ref{fig:spherex} follows a succession of infrared all-sky survey
programs, starting with the
InfraRed Astronomical Satellite (IRAS) \citep{Neugebauer84}, the Diffuse InfraRed Background Experiment
(DIRBE) on the COBE satellite \citep{Silverberg93}, 2MASS \citep{Skrutskie06}, AKARI
\citep{Murakami07}, and the Wide-field Infrared Survey Explorer (WISE) \citep{Wright10}, all of which 
mapped the full sky in broad photometric bands.  SPHEREx combines large-format sensitive infrared
detector arrays, an entirely passive cooling design, a wide-field off-axis telescope, and the novel
use of Linear Variable Filter (LVF) spectrometers to provide the first all-sky near-infrared
spectroscopic survey.  SPHEREx will produce 4 complete all-sky spectral maps over its 2-year
baseline mission.

We summarize SPHEREx core science in \S \ref{sec:science} and the mission requirements
to obtain these measurements in \S \ref{sec:requirements}.  The design of the mission, spacecraft, and
instrument are given in \S \ref{sec:implementation}.  We describe the on-orbit performance
and early observations in \S \ref{sec:onorbit} and conclude in \S \ref{sec:conclusion}.

\section{Science Overview}
\label{sec:science}

The design of SPHEREx flows from its 3 primary science objectives, constraining inflationary
non-Gaussianity, exploring the history of galaxy formation by studying the anisotropy of the
extragalactic background light (EBL), and surveying the abundance of water and other biogenic molecules
in the form of ices in the interstellar medium.  SPHEREx creates a unique all-sky spectral survey
to carry out these science objectives, enabling a diversity of investigations by the wider
astronomical community.  The SPHEREx science team will also curate 3 legacy catalogs with spectra
of solar system objects, exoplanet host stars and low-mass stars, and galaxy clusters.

\subsection{Inflationary Cosmology}
\label{ssec:ic}

SPHEREx tests the inflationary paradigm in modern cosmology, which was proposed to explain the large-scale
homogeneity and geometric flatness of the universe \citep{guth81, linde82, albrecht82, starobinsky80}.
The basic predictions of inflation have been
verified in increasingly precise measurements \citep{planck_VI_20}, specifically a near-flat spatial
geometry on large scales with initial density perturbations that are nearly Gaussian, phase-synchronous,
and nearly, but not precisely, scale-invariant.  The physical process driving inflation, however,
remains elusive.

Measures of cosmic non-Gaussianity are crucial for understanding the physics of the early Universe,
particularly the nature of primordial fluctuations generated during inflation. While standard inflationary
models predict the initial cosmic density fluctuations to be nearly Gaussian, deviations from Gaussianity
can reveal the presence of physical processes, such as interactions among fields during inflation or
cosmic defects.  Detecting or constraining non-Gaussianity thus provides valuable insights into the
fundamental mechanisms that shaped the Universe's structure and helps discriminate between competing
cosmological models.

\begin{figure}
\hspace{0.0cm}
    \includegraphics[trim={0cm 0cm 0cm 0cm}, clip, width=1.0\linewidth]{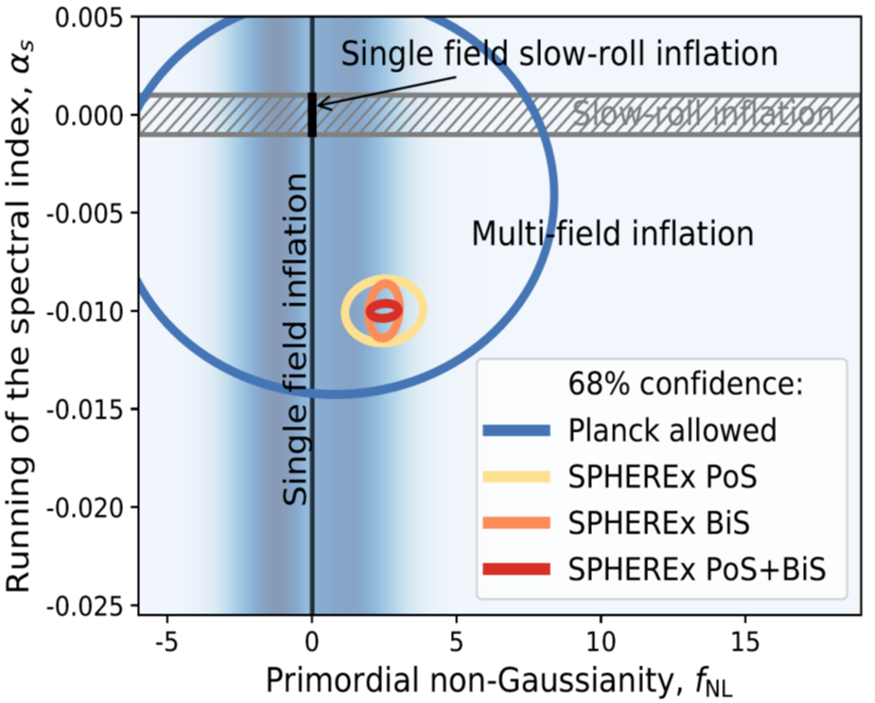}
    \caption{SPHEREx aims to establish strong constraints on non-Gaussianity $f_{NL}$ and running of the spectral index $\alpha_s$ in order to discriminate between broad classes of inflation models.  The plot identifies families of models, where the blue-shaded region illustrates the distribution of plausible multi-field models \citep{deputter17} with $|f_{NL}| \sim 1$, while single-field models with $|f_{NL}| < 10^{-2}$ fall along the y-axis.  Slow-roll models are shown in the hatched region with $|\alpha_s| < 10^{-3}$.  The current range of allowed parameters from Planck CMB data is shown by the blue ellipse.  SPHEREx's accuracy is shown for the power spectrum (PoS, yellow), the bispectrum (BiS, orange) and the combination (red).  While the centers of the SPHEREx ellipses are unknown, we place them at a location that would rule out single-field inflation for illustration.}
    \label{fig:fnl}
\end{figure}

Observations of non-Gaussianity in the primordial fluctuations, quantified by
the $f_{NL}$ parameters, describe the departure from a Gaussian bell curve \citep{komatsu05}, and
provide a unique test between multi-field and single-field models.  Inflation models driven by multiple
fields generally predict a local non-Gaussianity $|f_{NL}^{local}| \sim 1$ while simpler models with a single
field generally predict $|f_{NL}^{local}| < 10^{-2}$ \citep{alvarez14,dePutter:2016trg} (See Figure \ref{fig:fnl}).
For simplicity, we will write  $f_{NL}$ for $f_{NL}^{local}$ from now on. 

CMB-based measurements currently constrain $f_{NL} = -0.9 \pm 5.1$ \citep{planck_IX_20}, close to the
limits set by cosmic variance \citep{baumann09}.  Because large-scale structure galaxy surveys map the universe
in 3 dimensions, they access many more modes than a 2-dimensional CMB map, and can thus probe non-Gaussianity
to much lower levels than are possible with CMB measurements.  Current $f_{NL}$ constraints from galaxy
surveys are now starting to approach the accuracy of the CMB \citep{Chaussidon_2025}.

SPHEREx probes $f_{NL}$ using the power spectrum and bispectrum, the Fourier transforms of the 2-point and
3-point galaxy correlation function, respectively.  The relative sensitivity of these two measures is similar. 
Combined, they provide a powerful internal consistency test, because the two statistical measures are quite
different.  For the power spectrum, the effect of $f_{NL}$ is most prominent on large spatial
scales because of the scale dependent galaxy bias induced by $f_{NL}$ \citep{Dalal:2007cu}.  Its accuracy is
driven by maximizing the number of redshifts, which can have low accuracy
$\sigma_z = \Delta z / (1+z) < 0.2$, over a large cosmological volume.  
The bispectrum is driven by having multiple galaxy samples with smaller high-accuracy redshifts up
to $\sigma_z = \Delta z / (1+z) = 0.003$ to measure squeezed triangles with legs having a large spatial 
wavenumber $k$ \citep{Heinrich24}.

In addition to constraints on $f_{NL}$, we expect SPHEREx clustering measurements, with several hundred million
galaxies over the full sky, to lead to new insights on the properties of dark energy, the curvature of the Universe,
the mass of neutrinos \citep{Dore14}, and so-called general relativistic effects \citep{Wen:2024}.

\subsection{History of Galaxy Formation}
\label{ssec:gf}

Measures of the diffuse EBL are important because they capture the cumulative emission from all sources – both
resolved and unresolved – across cosmic history. This includes light from the earliest stars and galaxies,
diffuse intergalactic processes, and potentially new astrophysical phenomena. By studying the cosmic background
light, astronomers gain insight into star formation, galaxy evolution, and the total energy budget of the Universe,
including contributions that may be missed in classical point-source surveys. These measurements thus provide
a crucial census of cosmic structure and its evolution over time.

SPHEREx probes the history of galaxy formation by studying the spatial anisotropy of the EBL, the aggregate
glow of which contains emission from all sources integrated over redshift.  By mapping structures from degree
to 6\arcsec ~angular scales, SPHEREx accesses signals from linear (2-halo) clustering, non-linear
(1-halo) clustering, and Poisson (shot noise) fluctuations (e.g. \cite{Asgari23}, \cite{kovetz17}, \cite{COORAY02}).
Linear clustering refers to scales between dark matter halos that can be traced back to primordial fluctuations,
while non-linear clustering refers to smaller scales within a halo where the non-linear effects of gravity
and star formation are important.  Combined, these provide measures of the total light production, the brightness
of clustered galaxies and intrahalo light (IHL) associated with stripped stars in dark matter halos, and the
ensemble properties of galaxies above a selected brightness.

On large spatial scales, galaxies trace dark matter over-densities through linear clustering.  The linear
clustering signal is of particular scientific interest because it provides an integrated
measure of luminosity $b_I dI/dz$, where $b_I$ is the bias of emitting sources and $dI/dz$ is the total
light emitted at a specific redshift from all sources, including individually faint galaxies and diffuse
components that may be missed in galaxy-counting surveys.  SPHEREx has two ways to extract the redshift
history of emission.  Because each wave band is sensitive to a different range of redshifts, the
cross-spectrum between two bands partially correlates and provides information on the history of light
production.  The numerous cross-spectra thus enable fitting a suite of parameters to describe various
emission components and their luminosity history \citep{feng19}.  An alternative approach is to cross-correlate
SPHEREx maps with galaxy redshifts \citep{cheng22}, which provides an unambiguous redshift separation, but
is limited to available redshift surveys.

\begin{figure}
\hspace{0.0cm}
    \includegraphics[trim={0cm 0cm 0cm 0cm}, clip, width=1.0\linewidth]{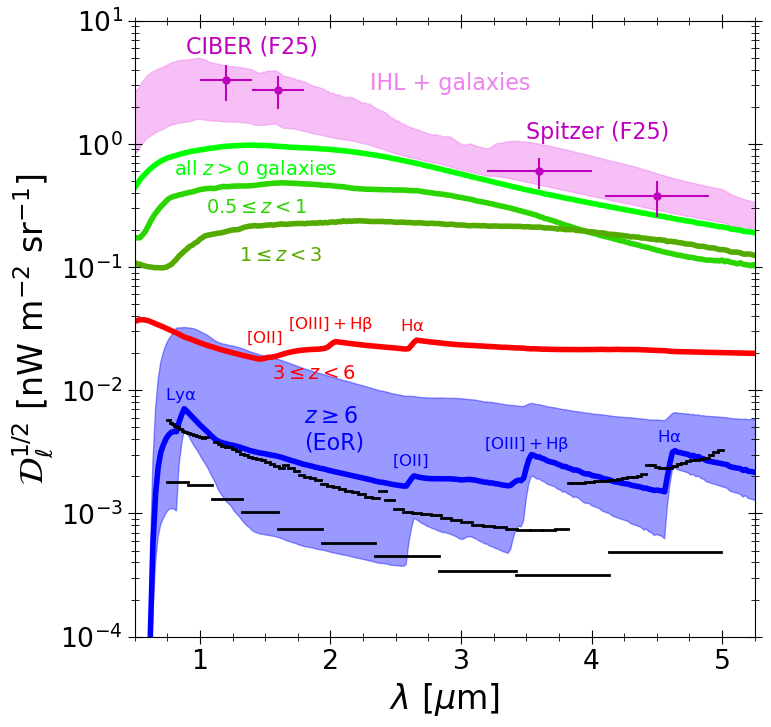}
    \caption{{\bf Color of EBL fluctuations on large scales, $\ell \simeq 10^3$.} Measurements from CIBER and Spitzer from \citet{feder25} (magenta points), all using the same mask of $\left\{\rm{J}=16.9,\rm{H}=16.9,\rm{IRAC}1=17.8 \right\}$, clearly exceed expectations from known galaxy populations fainter than the masking threshold (light green). Simple IHL models that adopt a linear relationship between the IHL fraction and dark matter halo mass (magenta band) can provide fluctuations at a level comparable to these available measurements, as suggested by previous work \citep{cooray12,zemcov14}. The sensitivity levels along the bottom (black lines) indicate SPHEREx's statistical sensitivity in the North Deep Field in a single $\ell$ bin $500 \leq \ell < 2000$ using idealized Knox errors, both in individual spectral channels (upper curves in black) and after binning into ten broad spectral bands (lower flat lines in black). Note that this $\ell$ bin, chosen to match the angular scale of linear galaxy clustering, is slightly wider than the measurements of \citet{feder25} at $\ell \sim$ 1300. Clearly, SPHEREx has sensitivity to galaxy populations across a broad range of redshift sub-intervals, including the EoR, as indicated by the annotated colored curves (model predictions from Mirocha et al., in prep.). Many spectral features may be visible in the EoR component, starting with Ly-$\alpha$ and the Lyman break at the blue edge of the wavelength range, and extending through strong rest-frame optical lines $[\rm{O}\textsc{II}]$, $[\rm{O}\textsc{III}]+\rm{H}\beta$, and finally $\rm{H}\alpha$ near the red edge of the SPHEREx band.}
    \label{fig:ebl}
\end{figure}

The SPHEREx survey (see \S \ref{ssec:surveydesign}) is designed with two $\sim$100 square degree deep field
regions (see Figure \ref{fig:deepfields}) that are ideal for these intensity mapping studies.  The fields are
located near the ecliptic poles, where the observing coverage naturally peaks due to SPHEREx's polar orbit.  The
coverage depth of the surveys is designed to provide sufficient statistical sensitivity to probe emission from
the epoch of reionization (EoR) at $z \ge 6$, with its characteristic Ly-$\alpha$ and Ly-cutoff features, in
the linear clustering signal (see Figure \ref{fig:ebl}).  The high-sensitivity and repetitive observations
in deep regions are also ideal for testing sources of potential systematic error (see \S \ref{ssec:systematics}).

\subsection{Interstellar Ices}
\label{ssec:ii}
Key biogenic molecules, such as water (H$_2$O), carbon dioxide (CO$_2$), carbon monoxide (CO) and methanol
(CH$_3$OH), are locked in ices \citep{boogert11, boogert13, oberg11} on the mantles of interstellar dust grains,
in amounts predicted in some cases to far exceed those in the gas phase \citep{hollenbach09}.  Ices play a
major role in forming planetesimals within disks, and are a key source of water and organic molecules for
newly forming planets.  Grain mantles are the sites of active surface chemistry \citep{tielens82, cuppen09}, and thus
one might expect ice composition to depend on their environment.  The limited number of existing ice spectra is
insufficient to determine whether the ices in protoplanetary disks, including our early Solar System, are mainly
inherited from their progenitor clouds \citep{dhooghe17}, or whether ice composition evolves over time \citep{aikawa12}.

\begin{figure}
\hspace{0.0cm}
    \includegraphics[width=1.0\linewidth]{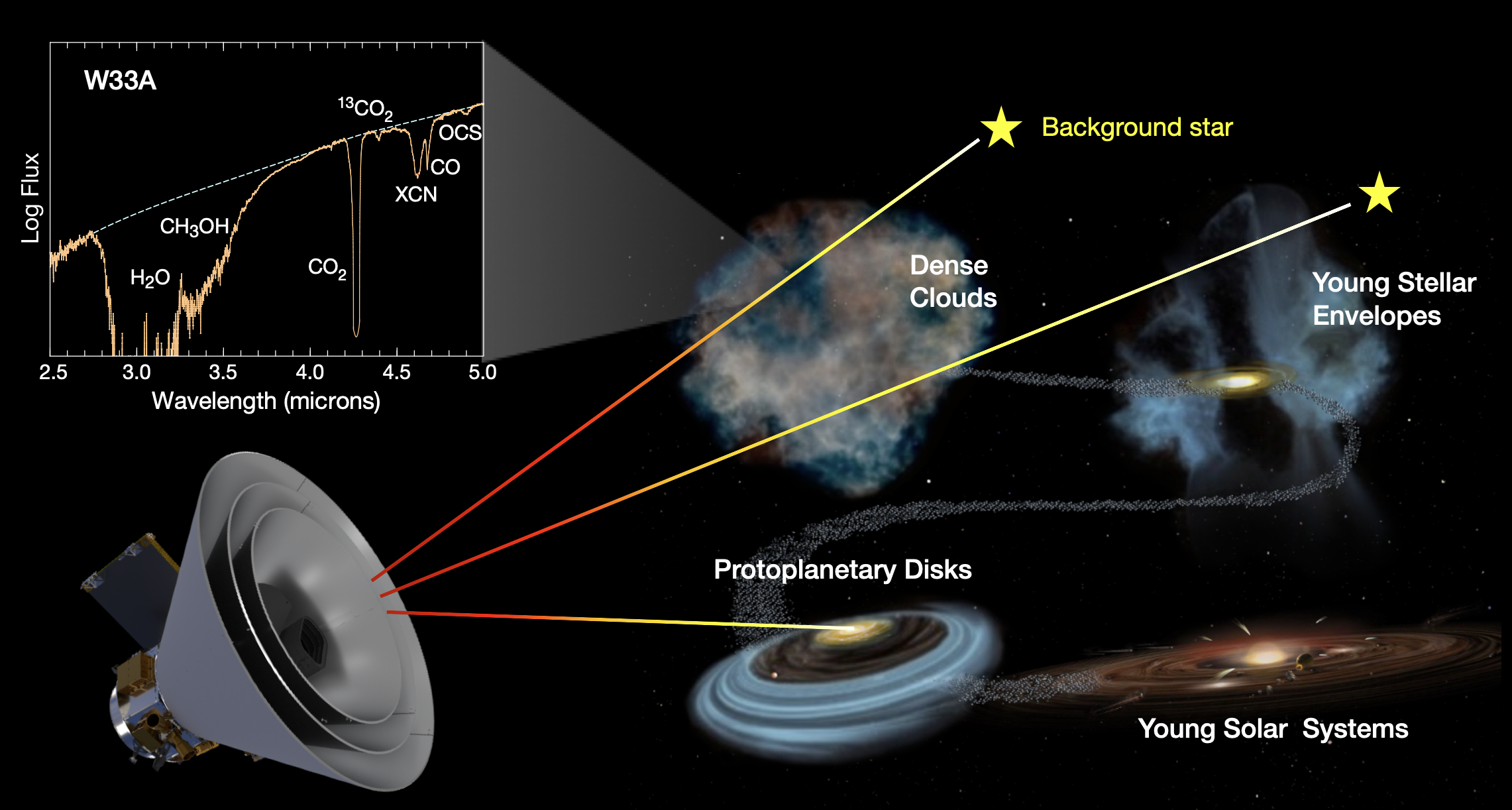}
    \caption{SPHEREx measures the abundances and properties of ice species by taking absorption spectra toward background stars and protostars.  This comprehensive survey toward $\sim$10 million targets spans the evolutionary stages of star and planet formation from dense molecular clouds to young solar systems with planetary disks.}
    \label{fig:ice_phases}
\end{figure}

SPHEREx addresses these questions by conducting a comprehensive survey of ices to better understand their abundances,
compositions, and how they evolve during the early phases of star and planet formation.  We selected a list of
more than 9 million targets based on 2MASS and WISE 3.4 and 4.6~\micron\ catalogs using three criteria:  1) a
line-of-sight extinction $A_V > 2$; 2) a WISE 4.6 $\mu$m flux brighter than $\sim$15 AB mag to ensure a high
signal-to-noise ratio in each SPHEREx spectral channel; and 3) targets that are separated from any neighboring
source contributing more than 1 \% of their flux by at least 7\arcsec.  The SPHEREx ices target list \citep{Ashby23}
will greatly expand -- by more than a factor of 10,000 -- the database of ice absorption spectra previously obtained
by ISO, AKARI, Spitzer and JWST, significantly advancing our understanding of how ices form and evolve as gas and
dust transition from diffuse to dense interstellar clouds, to circumstellar envelopes, and ultimately to
protoplanetary disks (\cite{Melnick2025}; see Figure \ref{fig:ice_phases}).

\subsection{All-Sky Spectral Survey}
\label{ssec:ass}

All-sky surveys (e.g. IRAS, COBE, AKARI, WISE) play a major role in advancing the field of astronomy, producing
versatile legacy archives that prove valuable for decades.  SPHEREx extends this tradition by providing
the first all-sky near-infrared spectral survey.  SPHEREx's rich spectral database supports numerous
scientific investigations \citep{dore16}.  These range from studies of main sequence, low-mass, and
evolved stars, to spectral catalogs of galaxies, to surveys of solar system objects, to observations
of clusters, to Galactic maps of line and PAH emission features.  The all-sky survey complements
recent and forthcoming astrophysics satellites \citep{dore18}, including JWST, TESS, Gaia, Euclid, eROSITA
and Roman.

\begin{figure}
\hspace{0.0cm}
    \includegraphics[trim={0cm 0cm 0cm 0cm}, clip, width=1.0\linewidth]{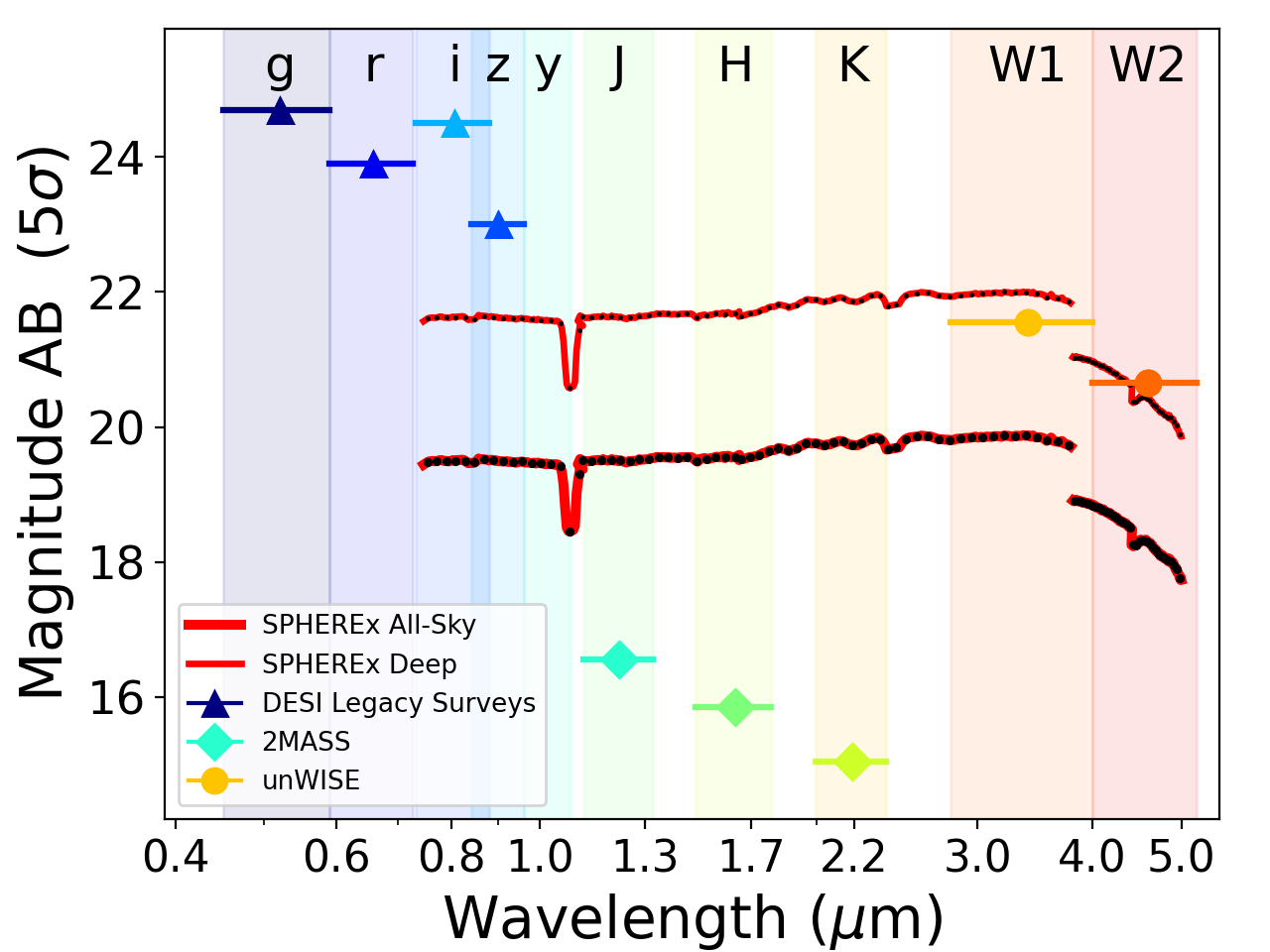}
    \caption{All-sky and deep field point source sensitivity derived from flight data (see \S \ref{ssec:sensitivity} and Figure \ref{fig:sensitivity}) as a function of wavelength. The thick red curve corresponds to the all-sky sensitivity, while the thin red curve corresponds to the deep field sensitivity, both after 2 years of observations. The black dots represent the sensitivity in the 102 independent wavelength channels. For reference, we add the depth of unWISE \citep{Lang14,Meisner17,Meisner19,Meisner22}, 2MASS \citep{Skrutskie06} and the DESI Legacy Imaging Surveys catalog \citep{Dey19}. Note that co-adding the SPHEREx spectral channels within the broad W1 and W2 bands gives an all-sky sensitivity that is very close to that of unWISE.}
    \label{fig:PSS_all-sky}
\end{figure}

SPHEREx also provides a valuable dataset for the upcoming generation of time-domain studies.
In the all-sky survey, SPHEREx measurements can serve as a reference every 6 months.  Note that it
typically takes 1-2 weeks to obtain a full 102-channel spectrum of any particular target in the all-sky survey.
In the SPHEREx deep fields with
50-100 times higher redundancy (see Figure \ref{fig:deepfields}), observations are much more frequent,
with a cadence that depends in detail on the location of the source.  As shown in Figure~\ref{fig:PSS_all-sky}, SPHEREx's
sensitivity extends deeper optical imaging surveys into the infrared, and provides spectral follow-up
for sources detected in WISE and 2MASS broadband photometry.

\subsection{Legacy Catalogs}
\label{ssec:lc}
In addition to data products that support the 3 core science themes (see \S \ref{ssec:ic} - \S \ref{ssec:ii}),
the SPHEREx science team is curating 3 spectral catalogs that benefit from the team's specialized knowledge
of the instrument:
asteroids and comets; planet-bearing and low-mass stars; and clusters of galaxies.  The solar system objects
catalog of asteroids and comets consists of objects with known, but changing positions.  \cite{ivezic22} estimates
that SPHEREx will deliver meaningful flux measurements for about 100,000 asteroids, of which $\sim$10,000 objects
will have high-quality spectra.  Because the spectral observations are distributed over time, care must be
taken to account for the effects of rotation and changing distance.  The wavelengths span from scattered
sunlight to thermal emission and include characteristic features of olivine, pyroxene, hydroxyl, water
ice, and organics, and contain information on thermal properties that complement broadband photometry with
the upcoming NEO Surveyor satellite \citep{mainzer23}.

SPHEREx will provide spectra to supplement external photometry from Gaia, GALEX, 2MASS and WISE for over 600,000
main sequence stars surveyed for exoplanets by the transit missions Kepler, K2 and TESS.  These stellar SEDs will
help determine best-fit effective temperatures, surface gravities, and [Fe/H] ratios to percent or sub-percent
precision.  When combined with Gaia astrometry, we can derive bolometric luminosities, proper motions, stellar
radii and extinctions.  This combination of data significantly improves planetary radius estimates for
many transiting planets, by reducing the error from uncertainty in the stellar radius, which can otherwise
dominate, especially for cool stars \citep{stevens18}.  SPHEREx will also compile a spectral atlas of
~10,000 low-mass stars, from late M dwarfs and all accessible brown dwarfs down to the coolest Y dwarfs.
Low-resolution near-infrared spectroscopy is crucial for studying these cool stellar objects, which can
have rich spectra due to their complex atmospheres.  The observations will use Gaia astrometry to account for
significant proper motions.

Finally, SPHEREx will produce a spectral catalog of $\sim$100,000 galaxy clusters.  Galaxy clusters
are large, gravitationally bound structures, consisting of constituent galaxies, dark matter, and hot
intra-cluster gas visible at X-ray wavelengths and through the millimeter-wave Sunyaev-Zel'Dovich effect.
Clusters form late in the $\Lambda$CDM hierarchical model of galaxy formation and are thus interesting
subjects for cosmology, as well as for elucidating star formation and feedback processes in galaxies.
Using photometry tools optimized for crowded fields, the science team will determine fluxes
and estimated redshifts of member galaxies.  The SPHEREx data will be assembled together with available
optical, X-ray and millimeter-wave observations.  For selected clusters the data may allow estimates
of the stellar mass and ages of cluster galaxies, as well as emission line fluxes and equivalent widths.

\section{Science to Requirements}
\label{sec:requirements}

SPHEREx is designed to fulfill the science requirements of its three primary scientific objectives.
In short, the inflationary cosmology theme
is driven by having sufficient high-accuracy and low-accuracy redshifts to constrain the non-Gaussianity
parameter $f_{NL}$ through the bispectrum \citep{Heinrich24} and power spectrum, respectively.  These
in turn set requirements for spectral resolving power and point source sensitivity in bands 1-4 (see Table
\ref{table:spectro}).  The history of galaxy formation theme can carry out its science with
the same resolving power and sensitivity, but must control noise and systematic errors on degree angular
scales in order to accurately map extended cosmological structures.  The interstellar ices science theme
studies bright Galactic sources, but requires higher resolving power, with wavelength coverage out to 5 $\mu m$
to spectrally separate ice absorption features.  We describe the flow of engineering requirements from science
in abbreviated form, but note that extensive simulations were carried out to justify these choices
during the mission development. 

\subsection{Spectral Coverage and Resolving Power}
\label{ssec:specres}
The four short-wavelength bands were designed for the cosmological redshift survey, with
resolving powers of $\lambda / \Delta \lambda \sim 40$ (see Table \ref{table:spectro}).  We carried
out simulations \citep{Stickley16} of SPHEREx observations on the COSMOS field, using realistic
galaxy spectra and including the effects of spectral confusion from faint galaxies below SPHEREx's
sensitivity.  These studies indicated that we could obtain redshifts with accuracy up to
$\sigma _z = \Delta z / (1+z) \leq 0.003$, required for the bispectrum analysis, using only continuum
features.  The most recent study of SPHEREx redshifts \citep{Feder24} confirms our choice of resolving
power is appropriate, and indicates we could enhance the number of high-accuracy redshifts by incorporating
emission lines.  The galaxy formation theme can be accomplished with the wavelength and spectral resolution
choices for cosmology.

The two long-wavelength bands were designed to measure interstellar ices, with resolving powers of
$\lambda / \Delta \lambda$ = 110 and 130 (see Table \ref{table:spectro}).  The constraints of the survey
design require a fixed number of spectral channels per band (see \S 4.3.3 and Figure \ref{fig:bandc}),
so the spectral range is inversely related to the resolving power.  Our chosen criterion was to first cover
wavelengths up to 5~\micron\ to establish the underlying continuum, and then to measure the equivalent
width of the CO, CO$_2$, CH$_3$OH, OCS, and XCN features, cleanly separating the 4.62 $\mu$m XCN  and
4.67 $\mu$m CO ice features.  The broad H$_2$O ice feature at 3~\micron\ is easily resolved.  However,
the chosen resolution is lower than needed to fully resolve all the ice features.  We simulated
observations from measured ice features over a range of extinction and stellar type of the
background star to substantiate these choices.

\subsection{Point Source Sensitivity}
The inflationary cosmology theme constrains inflationary non-Gaussianity through measurements of the
power spectrum and bispectrum with a large catalog of galaxies with a range of redshift accuracy \citep{Dore14}.
The number density of galaxies as a function of redshift, their clustering biases, and their redshift accuracy set
the measurement error on $f_{NL}$.  The forecasting methodology used for the power spectrum is described in
\cite{Dore14} and \cite{deputter17}. The forecasting methodology used for the bispectrum follows \cite{Heinrich24}
and includes an accurate handling of redshift errors, but excludes the non-Gaussian contribution to the covariance
matrix \citep{Biagetti:2021tua, Salvalaggio:2024} and the covariance between the power spectrum and bispectrum
\citep{dePutter:2018}. It also assumes the galaxy biases universality relation \citep{Dalal:2007cu,Barreira:2020ekm}.
All these effects, when included, might reduce the $f_{NL}$ sensitivity by tens of percent.  However, a full analysis
incorporating a multi-tracer analysis with optimal samples may substantially improve the measurement error on $f_{NL}$ from our initial forecast.

\begin{figure}
\hspace{0.0cm}
    \includegraphics[trim={0cm 0cm 0cm 0cm}, clip, width=1.0\linewidth]{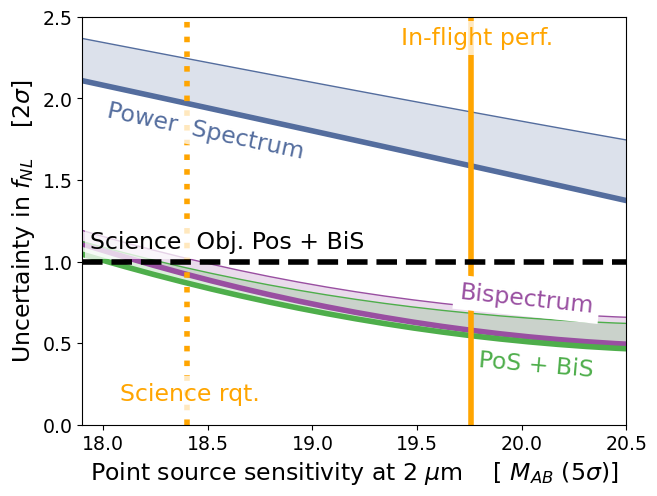}
    \caption{Translating accuracy on the $f_{NL}$ parameter to point source sensitivity referenced at 2 $\mu$m. These curves were obtained simulating SPHEREx all-sky survey photometry on a COSMOS-based field to obtain realistic number density of galaxies as a function of redshift. The galaxy number densities were then processed with our cosmological forecast machinery. The thick solid curves correspond to statistical error (power spectrum in blue, bispectrum in purple and power spectrum and bispectrum combined in green). The systematic error allocation of $\delta n/n < 0.2 \%$ rms per dex is shown by the shaded region bounded by a thin line.  The in-flight SPHEREx median point source sensitivity of 19.7 AB mag (see \S \ref{ssec:sensitivity} and Figure \ref{fig:sensitivity}) is given by the vertical line.}
    \label{fig:PSS_fnl}
\end{figure}

\cite{Feder24} studied the number
and accuracy of redshifts using a template fitting method, which strongly depends on point source
sensitivity in bands 1-4.  In order to achieve the accuracy $\Delta f_{NL} < 0.5$ needed to meaningfully
discriminate between single-field and multi-field inflation models, we set a requirement that the
point source sensitivity exceed 18.4 AB mag ($5\sigma$) per $\Delta\lambda_{SC}$ spectral channel
(see \S \ref{ssec:surveydesign}) in bands 1-4 once observations are complete (see Figure~\ref{fig:PSS_fnl}).

The galaxy formation theme can deeply probe for the faint EOR clustering anisotropy signal with
the surface brightness sensitivity that comes from the point source sensitivity requirement.  The
interstellar ices theme can extract a required 20,000 high-quality ice-absorption spectra, with a
signal-to-noise $>$ 100 per spectral channel, with a non-driving sensitivity requirement of 9 AB mag
($100\sigma$) in bands 5-6.

\subsection{Spatial Resolution}
\label{ssec:spatres}
The requirement for spatial resolution flows from the goal of obtaining redshifts for inflationary
cosmology science.  Given prior knowledge of the source positions, the number and accuracy of the redshifts
SPHEREx measures are driven by instrument point source sensitivity.  Therefore, the requirement on spatial
resolution is to select the minimum pixel size that does not degrade point source sensitivity in the all-sky
survey.  We give a detailed calculation for this optimization in the appendix, with the constraints of a fixed
number of pixels and fixed observation time, which concludes that the choice of 6\arcsec ~pixels is optimal.

Using the simulation tools described in \cite{Stickley16}, we studied photometry on point sources in the
presence of surrounding stars and galaxies below the SPHEREx detection level with variously sized pixels and PSFs.
These calculations confirmed that photometry is limited by instrumental (photon noise dominated) sensitivity for
the all-sky survey, and the effects of confusion are small.  Consequently, reducing either the PSF or the pixel
size has little impact on the number of measured redshifts.  Indeed, we observed that a smaller PSF with a fixed
6\farcs2 pixel size improved photometric sensitivity, though very modestly, by slightly reducing the effective
number of pixels $N_{eff}$ observing a point source.

Spectral confusion adds wavelength-coherent variations that depend on the pixel scale and the depth of sources
extracted for photometry \citep{Huai25}.  For the all-sky survey, 6\farcs2 pixels limit the
density of targeted sources due to beam overlap increasing noise, but spectral confusion itself is small
compared with instrument noise.  At the higher sensitivity in the deep fields, spectral confusion may
exceed instrument noise.  For the galaxy formation science theme in \S \ref{ssec:gf}, the requirement
is to mask or otherwise remove point sources from the map.  For this science, 6\farcs2 pixels suffice,
because improved spatial resolution only helps remove the faintest sources.  Finally, \citet{Huai25} quantify the
photometric depth and galaxy redshifts that can be reliably extracted from the deep field due to the effect of
spectral confusion.

\subsection{Systematic Errors}
\label{ssec:systematics}

Systematic errors must be mitigated to levels that meet the three core science objectives.  SPHEREx uses
a tiered strategy to:  1) control systematic errors in the instrument design; 2) assess systematics
through astrophysical data; and 3) use in-flight measurements to further mitigate systematics in
data analysis, if necessary.  We confine our attention here to an abbreviated discussion of systematic
errors arising from the instrument, though additional effects from astrophysical phenomena and external
data sets will also need to be considered during science analysis.  For the cosmology theme, we budget
systematic errors into a total allocation
for spurious clustering which totals to $\delta n/n < 0.2 \%$ rms per dex (see Figure \ref{fig:PSS_fnl}).
The most important instrumental effects that enter are gain stability, noise bias and photometric
accuracy.  In order to accurately
map degree-scale structures, the galaxy formation theme must control errors arising from stray light
and dark current, ideally below the level of statistical uncertainty.  Finally, the interstellar ices
theme requires accurate channel-to-channel photometry in order to recover the depth of ice absorption
features.  We describe the general approach for each of these effects, but the final quantification
must await a full analysis of flight data.

\subsubsection{Gain Stability}
We intend to correct gain variations to $< 1$ \% over 2 years for inflationary cosmology science.  The
instrument controls gain stability by addressing the detectors with stabilized biases and references that are
on thermally regulated oven-controlled regions in the warm electronics circuit boards, and by thermally
regulating the H2RG detectors.  The WISE mission shows remarkable stability with its H1RG arrays
\citep{cutri18}, with $<0.5\%$ variation in amplitude over 8 years after correcting for drifts in
the uncontrolled detector temperatures.  SPHEREx will monitor gain stability by measuring the relative
flux of $\sim$100,000 bright stars in the deep fields.  The deep fields are generally observed every
orbit, and the stars provide ample statistical sensitivity to track relative gain changes in every
readout channel.

\subsubsection{Noise Bias}
If uncorrected, variations in sensitivity on the sky lead to spurious clustering due to redshift non-uniformity.
Because SPHEREx is photon noise limited, sensitivity variations are driven by spatial variations in sky
brightness, primarily due to variations in Zodiacal light.  We plan to mitigate the amplitude of the
raw effect on spurious clustering \citep{Dore14} by a factor of $\sim$20 by injecting artificial sources
into the data pipeline \citep{huff14, suchyta16,Everett:2020}.  Since the artificial sources are observed with noise
realizations of real data, they accurately track the variations in redshift number density due to variations
in sensitivity.  We also model the detector noise with photocurrent \citep{Nguyen25}, and record daily
image pairs taken of the same field to provide an accurate in-flight assessment of the noise behavior
per pixel.

\subsubsection{Photometric Accuracy}
\label{sssec:photometric}
In order to control systematics in the optimal photometry (see \S \ref{ssec:spatres}) of
galaxy redshifts and interstellar ice features, SPHEREx requires a careful determination of the
instrument PSF, a precise determination of the flat-field response of pixels over the focal plane,
and an accurate calibration, both between spectral channels and overall for comparison with other datasets.
The PSF is measured in small regions of the focal plane by sub-pixel stacking star images with
accurate astrometry, and deconvolving the pixel response function.  The flat-field response matrix
connects the gain of pixels within a spectral channel to discrete locations measured with calibration
stars.  We developed a method that operates on an ensemble of images taken over the mission at varying
Zodiacal brightness levels.  The response of each pixel is regressed against a tracer of the Zodiacal
light.  We then divide this Zodiacal
template image by the overall spectral response to Zodiacal light with wavelength to obtain the
flat-field matrix.  The regression offsets also provide an estimate of the dark current per pixel.
We apply small corrections to this method to remove bright Galactic regions, account for variations
in the pixel size due to optical distortion, and remove the brightness gradient with solar elongation.
The top-level calibration target is 3.0 \% (2.0 \%) for overall (channel-to-channel) accuracy.
Of this, we allocate 2.0 \% (1.5 \%) to uncertainties in the calibrators, to be added in quadrature
with 2.1 \% (1.1 \%) allocated to the instrument and 0.7 \% allocated to the combined errors from
the PSF and flat-field determinations.

\subsubsection{Stray Light}
SPHEREx uses an off-axis telescope design to control stray light on scales ranging from the 6\arcsec ~PSF
core to tens of degrees.  An extended PSF produces `halos' that are problematic
around bright sources.  The design reduces the extended PSF by controlling mirror roughness,
particulate contamination, and multiple reflections between the LVFs and detectors.  The extension
is well measured in flight data and can be mitigated by a combination of masking and numerical
subtraction.  The response to bright stars outside the field of view is controlled by a series
of baffles around the mirrors and detectors.  Finally, far-angle response to the Earth and Galactic
plane are controlled by the telescope design.  We have carried out a series of in-flight measurements
to assess the level of systematic error that results in the deep field maps used in the galaxy
formation science theme.  We will further check susceptibility by comparing subsets of the deep maps,
made under different stray light conditions, for statistical consistency.  A detailed analysis of the
design and the in-flight measurements will be presented in a future publication \citep{Dowell25}.

\subsubsection{Dark Current}
Detector dark current produces a systematic error for galaxy formation science by leaving a residual
pattern in the deep field maps.  The dark current levels reported in Table \ref{table:fpas} were
measured in the laboratory under flight-like conditions, using the flight cadence and regular resets
in 32-channel output mode.  We simulated deep field mosaics from these lab dark current images,
and find a residual signal in a synthesized broad $\sim$20 \% band at 1.2 $\mu$m at the level of
$D_{\ell}^{1/2} \simeq 7 ~{\rm pW m}^{-2}$ at an angular scale of $500 < \ell < 2000$.  To reach
our target level for this systematic, we plan to measure
and subtract dark current to a modest accuracy of $< 20$ \% using the offsets produced by the
Zodiacal flat-field estimator described in \S \ref{sssec:photometric}.  We have already produced
dark current estimates using this method on early flight data and find general agreement with
the laboratory measurements.  Once the deep fields are fully observed, we will construct null
tests with chosen combinations of images \citep{zemcov14} to quantify the level of systematic
error that arises from dark current and other mechanisms.

\subsubsection{Voxel Completeness}
\label{sssec:completeness}
The four surveys we planned in the baseline mission are necessary to build sensitivity, interleave
spectral sampling, identify time-variable sources, and test photometric reliability with independent
measurements.  We set a modest completeness of spectral-spatial elements ('voxels') to be
$> 90$ \% per survey.

\section{Implementation}
\label{sec:implementation}

SPHEREx is a NASA medium class Explorer (MIDEX) satellite, managed and operated by the Jet
Propulsion Laboratory (JPL) for the principal investigator, James J. Bock.  JPL
provided the payload thermal system, consisting of the photon shields, V-groove coolers, and bipod supports, and 
developed the focal plane mechanical assemblies.  British Aerospace (BAE) Systems (formerly
Ball Aerospace \& Technologies Corporation) provided the telescope and spacecraft.  The California
Institute of Technology (Caltech) provided readout electronics and the dichroic beamsplitter (DBS).  
Teledyne Imaging Systems provided the detector arrays and Viavi Solutions developed the LVFs and
coated the DBS.  Caltech characterized the integrated instrument for focus, spectral
response, and calibration.  BAE and JPL integrated the instrument, payload thermal system, and spacecraft into an 
assembled observatory, and carried out environmental testing prior to launch.  The Korea Astronomy and
Space Science Institute (KASI) provided a cryogenic vacuum chamber to test the instrument.  KASI currently
curates the solar system objects catalog, and contributes to scientific data analysis.  IPAC handles the data
processing and analysis through Level 3 (see \S \ref{ssec:pipeline}), and releases data products and
analysis tools to the public through IPAC's Infrared Science Archive (IRSA).  The basic specifications
of the SPHEREx observatory (see Figure \ref{fig:spherex}) are summarized in Table \ref{table:specs}.

\begin{table}[h]
    \centering
    \begin{tabular}{l|l}
        Parameter & Value \\
        \hline
         Launch Mass & 499.9 kg\\
         Outer Dimensions & $\phi$320 $\times$ 230 cm\\
         Solar Panel Size & 267 $\times$ 102 cm\\
         Power & 271 W\\
         Science Data Volume & 20 GB/day\\
         Orbit & Sun-synchronous\\
         Semi-Major Axis & 7037.4 km\\
         Orbital Inclination & 97.951 deg\\
         Orbital Ellipticity & 0.00034\\
         Longitude of Ascending Node & 27.06 deg\\
         Mean Local Time of Ascending Node & 6:00 AM\\
         Telescope & Off-axis 3-mirror free-form\\
         Effective Aperture & 20.0 cm\\
         Pixel scale & 6.2 arcseconds\\
         Detectors & 6 x H2RG\\
    \end{tabular}
    \caption{Table of basic parameters for the SPHEREx observatory}
    \label{table:specs}
\end{table}

\begin{table}[h]
\centering
    \begin{tabular}{c|c|c}
        Band & Wavelength Range & Resolving Power \\
        \hline
         B1 & 0.75 - 1.11 $\mu$m & 41\\
         B2 & 1.11 - 1.64 $\mu$m & 41\\
         B3 & 1.64 - 2.42 $\mu$m & 41\\
         B4 & 2.42 - 3.82 $\mu$m & 35\\
         B5 & 3.82 - 4.42 $\mu$m & 110\\
         B6 & 4.42 - 5.00 $\mu$m & 130\\                       
    \end{tabular}
    \caption{Properties of the LVF spectrometers.  Note the wavelength range is given for
    wavelengths that are present across the array.  There are small regions with further
    coverage on each detector that provide some spectral overlap between bands.}
    \label{table:spectro}
\end{table}

\subsection{Mission Design}
SPHEREx adopted a Sun-synchronous orbit for reasons of thermal performance, following
the examples of IRAS, COBE, AKARI and WISE.  To facilitate mapping the sky, we chose an orbital
inclination of $97.951 ^\circ$ that gives a precession rate of 360 degrees per year.  Based on
the achieved orbital parameters, we forecast that the current SPHEREx orbit would enable
continued survey observations for more than 15 years.

The observatory
conducts observations in a restricted sky region, keeping the Sun at least $91 ^\circ$ off the axis
of the photon shields.  The observatory collects data in 116.9 second
exposures while the observatory pointing is stable.  The exposures are generally close to the great
circle $90 ^\circ$ from the Sun.  However, we exploit the freedom to depart the great circle, within
the limits of the Sun, Earth, and Moon avoidance constraints, to cover the sky
efficiently.  The telescope boresight is tipped towards the solar panels by $8^\circ$, which is
needed to cover the full sky, especially at the ecliptic poles, while maintaining the $93.5 ^\circ$
Sun avoidance angle.  In contrast, (IRAS and WISE) kept the telescope bore sight aligned with the
spacecraft z-axis, and adopted a canted sun shield to allow a Sun avoidance angle less than 90 degrees.

\subsection{Survey Design}
\label{ssec:surveydesign}
The SPHEREx survey \citep{bryan25} is designed to cover the entire sky, observing every point on the sky
in each of its 102 spectral channels, at least twice per year.  We define spectral channels in intervals
given by $\Delta \lambda _{SC} = \int T(\lambda ) d\lambda$, where $T(\lambda )$ is the LVF transmission
normalized to unity at peak.  The required spectral range and resolving power set the minimum
number of spectral channels.  However, to simplify the LVF fabrication we fix the resolving power per band,
and to simplify the survey design we require that all the spectral channels subtend the same angular
size on the focal plane.  Our design thus has 17 spectral channels (see \S 4.3.3 and Figure \ref{fig:bandc})
for each of the 6 bands in Table \ref{table:spectro}.  In the survey design, we consider a spectral
channel to be measured when the observed wavelength falls within the geometrically-defined region.  Note
that each object will actually be observed at a unique set of wavelengths, and will typically have more
than 102 wavelength observations in each survey.  Because the interval of a spectral channel does
not Nyquist sample the spectral response of the LVF, we shift the targets by half a spectral channel, i.e.
$\frac{1}{2} \Delta \lambda _{SC}$, between the first and third, and second and fourth surveys.

\begin{figure*}
\hspace{-0.1cm}
    \includegraphics[trim={0cm 0cm 0cm 0.9cm}, clip, width=1.1\linewidth]{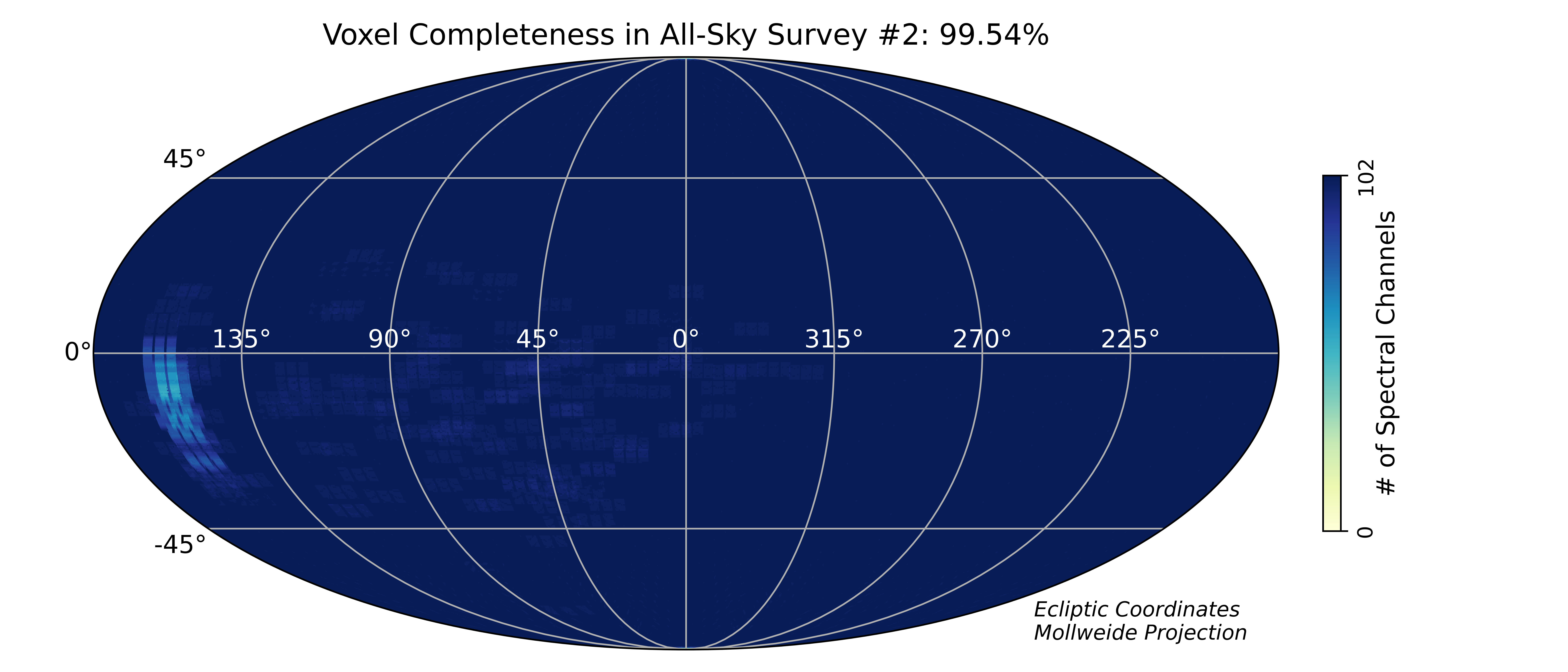}
    \caption{Simulated coverage of the second of the four all-sky surveys, shown as the number of spectral channels observed per 6\farcs2 sky pixel. This survey has the lowest voxel completeness of the four, with 99.54\% observations of all spectral channels over all sky pixels.  The region at the left with incomplete coverage is
    due to moon avoidance.}
    \label{fig:allsky}
\end{figure*}

\begin{figure}
\hspace{0.2cm}
    \includegraphics[trim={0cm 0cm 0cm 0cm}, clip, width=0.95\linewidth]{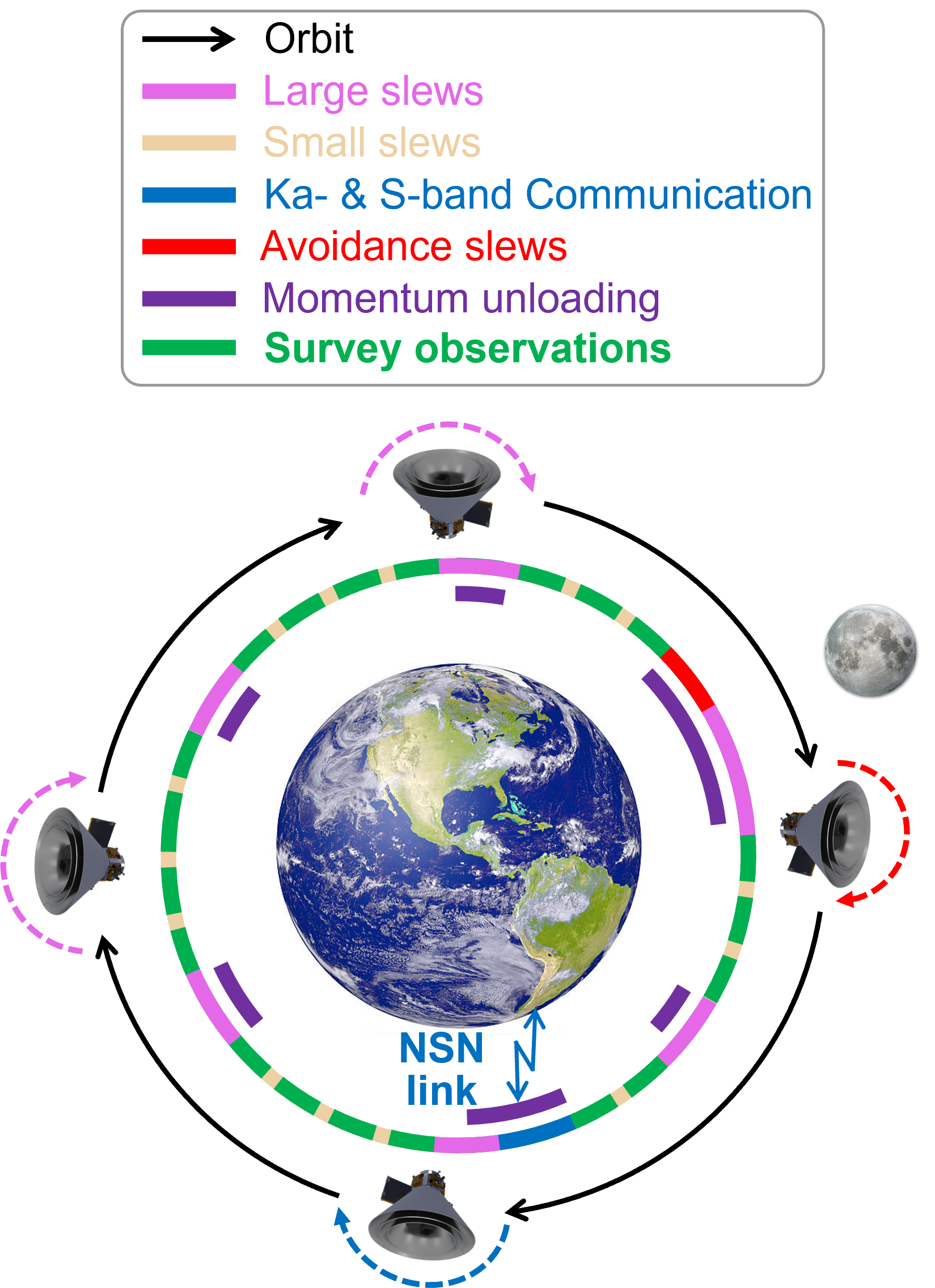}
    \caption{Schematic of activities during operations.  The spacecraft executes a series of pointed exposures over an orbit (green), separated either by small slews (tan) or large slews (pink) to maintain avoidance criteria to the Earth, Sun and Moon.  During periods where the Moon is visible, the spacecraft may make a larger slew (red) to avoid coming too close to the Moon.  Several times a day the satellite telemeters science data to the Near Space Network (NSN) (blue) and then returns to observing.  The spacecraft unloads momentum (purple) via magnetic torque rods during large slews.}
    \label{fig:orbit}
\end{figure}

To cover the sky, the observatory slews to a series of pre-programmed targets, maintaining
avoidance criteria to the Sun, Earth and Moon. The motion between pointings is generally along the
gradient direction of the LVFs (see Figure \ref{fig:bandc}).  The observations are typically executed
in a  series of 1 to 4 exposures, where an exposure produces a set of 6 spectral images, one per
detector.  Exposures in a set are separated by small slews of 11.8 arcminutes, chosen to match the
distance between spectral channels and to minimize time lost between exposures.  However, the motion
of the orbit around Earth eventually requires a large slew, with an angular length of up to $70^\circ$,
to maintain Earth avoidance (see Figure \ref{fig:orbit}).  The survey plan interleaves telemetry passes
with the observing sequence to ensure smooth operations.

The survey design does not count observations in the South Atlantic Anomaly (SAA) region towards completeness
(see \S \ref{sssec:completeness}), although the images are still taken and the data telemetered.
Finally, we allow for a small fraction of the images, 1 to 3 \% on average, to be flagged for potential reobservation
for reasons of data quality.  Presently we use a criterion based on the fraction of transient events, mainly
from energetic particles hitting the detectors, flagged during the exposure to reobserve the most degraded
images.  As we gain familiarity with the performance, and solar activity varies with the solar cycle, more
sophisticated criteria may be adopted in the future.

\subsubsection{Earth and Sun Avoidance}
\label{sssec:earthsun}
We set criteria for the survey to avoid the Sun and Earth throughout observations. Autonomous fault
protection onboard the satellite maintains more conservative avoidance angle criteria.  Here we give
the avoidance angles that we use with our Survey Planning Software that have margin to the fault
protection values.  The survey keeps the Sun $> 93.5 ^\circ$ off axis, so that sunshine never
illuminates the inner photon shield where heating and stray light response would be unacceptable.
As the spacecraft z-axis points away from local zenith, earthshine starts to illuminate the inner
photon shield for zenith angles larger than $25 ^\circ$, and specularly reflects to space.  As the
angle increases, earthshine will illuminate the top of the telescope baffle and the black radiator
panel that cools the MWIR detectors.  We originally defined the earth avoidance criteria to always
keep the zenith angle smaller than $39.3 ^\circ$. However, in-orbit measurements of the stray light
response to the Earth's limb (see \S \ref{ssec:straylight}) showed that the zenith angle could be increased.
Therefore, we adopted the criterion that the zenith angle should be $< 45^\circ$.  Furthermore, from
in-orbit observations, we learned that the detectors see strong shuttle glow emission
(see \S \ref{sssec:shuttleglow}) at 3 and 4.5 $\mu m$ when the observatory is tipped towards our
direction of flight (i.e. the ram direction).  To mitigate this, we set a criterion that the telescope
point $< 20^\circ$ from zenith in the ram direction. While this significantly reduces the amplitude
of shuttle glow, it is still present in the images at a low level.  Taken together, the allowable
pointing zone is now larger than the pre-launch estimate, improving our ability to survey the sky
efficiently without losing more time to large slews and other overheads.

\subsubsection{Moon Avoidance}
We avoid the moon by $> 32^\circ$ from the telescope boresight during observations, while maintaining
a less restrictive angle of $>13^\circ$ during slews.  

\subsubsection{South Atlantic Anomaly}
The survey design excludes a region associated with the SAA, accounting for 8.1 \% of the Earth's
surface area and modeled as a quadrilateral centered on -26$^\circ$/-45$^\circ$ lat/lon, 
where the transient rate on the detectors is significantly elevated.  The survey is designed to fully
cover the sky while excluding data obtained in the SAA.  However, images from the SAA region
are still telemetered and available for analysis by the science team.

\subsubsection{Deep Fields}
\begin{figure*}
\centering
    \includegraphics[width=1\linewidth]{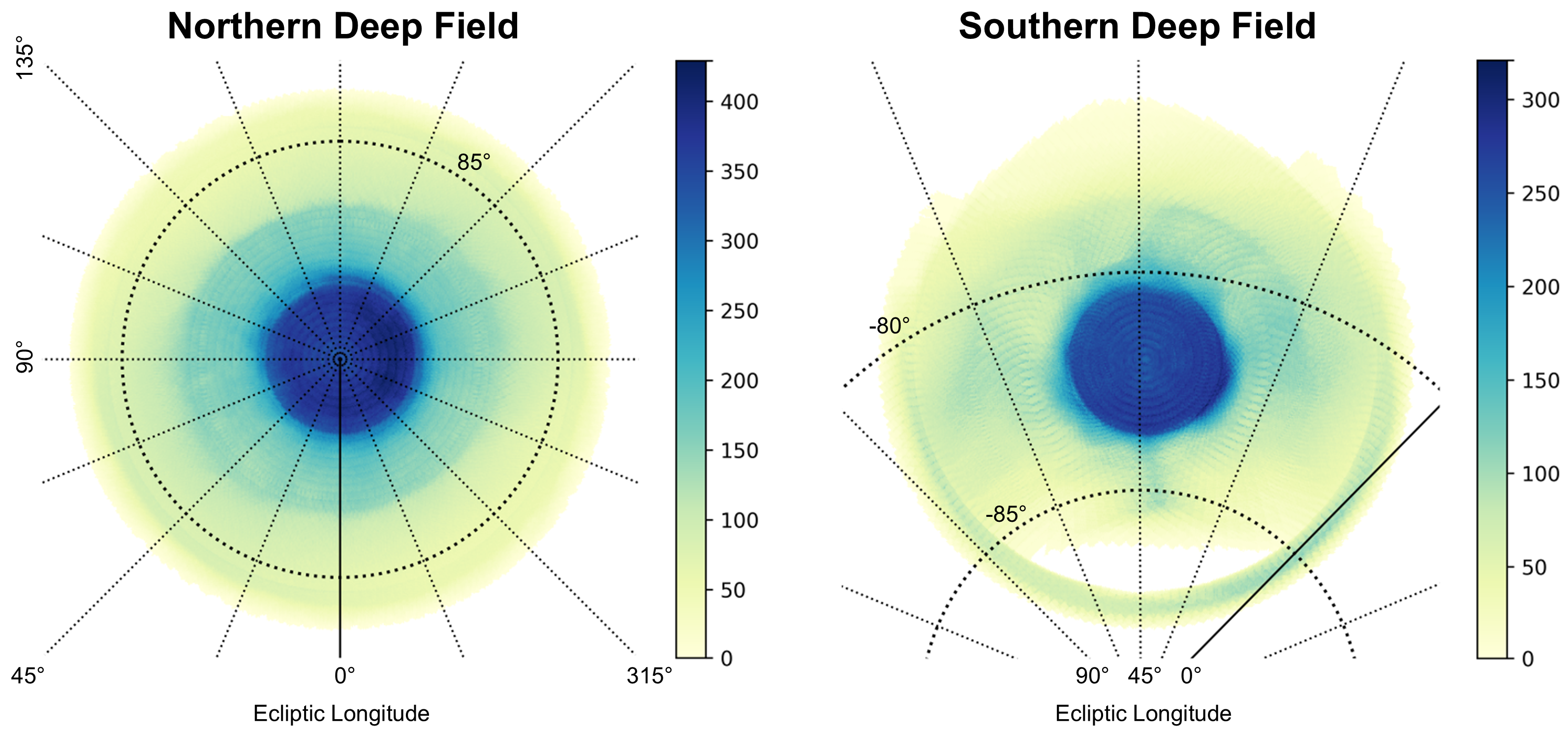}
    \caption{Simulated coverage of the SPHEREx deep fields, illustrated as the minimum number of observations per 6\farcs2 sky pixel across all 102 spectral channels after two years of planned observations.  The northern field is centered on the north ecliptic pole, while the southern field is offset from the south ecliptic pole and centered at ecliptic latitude $-82 ^\circ$ and ecliptic longitude $-44^\circ.8$ to avoid the Magellanic Clouds.  Ecliptic coordinates are overlaid for reference.}
    \label{fig:deepfields}
\end{figure*}

Each orbit, as the spacecraft points at target fields to cover the full sky, it also observes targets
in two deep survey fields located near the north and south ecliptic poles that were chosen for the
galaxy formation science theme.  These deep surveys are designed to map the smallest field possible,
each about 100 square degrees due to SPHEREx's large field of view, with complete spectral coverage.
As shown in Figure \ref{fig:deepfields}, the expected depth varies over each field, peaking at more
than 400 observations in every channel at the center and gradually falling off at the edges.  The
southern field is displaced from the ecliptic pole in order to avoid the Magellanic Clouds, although
this results in less favorable coverage compared with the northern field.

\subsubsection{Survey Cadence}
The survey proceeds by filling in observations near the great circle $90 ^\circ$ from the Sun in piecewise
fashion.  Over the course of each survey, a typical object on the sky will be observed in all of the
spectral channels within approximately two weeks.  Because the LVF wavelengths vary continuously over the
field of view, the exact central wavelengths will slightly vary, with a unique pattern for each object.
During the first year, we define survey 1 where the spectral gradient of the LVFs points toward the
ecliptic South pole, and survey 2 where the gradient points toward the North.  Thus after approximately
6 months the sky will be fully observed, with roughly half the observations in survey 1 and half in
survey 2.  After the baseline 2 year mission, the sky will be observed 4 times in all of the spectral
channels, and objects will have 2 Nyquist-sampled spectra.

\subsection{Instrument Design}
As a NASA Explorer satellite, SPHEREx is designed to carry out its scientific objectives with an
economical implementation.  We developed the observatory to carry out its survey using a spectrometer
without moving parts, based on an entirely passive cooling system and a single-string architecture throughout
the instrument and spacecraft.  Further discussion of the instrument design, testing, and performance can be
found in \citet{Korngut25}.

\subsubsection{Cooling System}
SPHEREx uses a multi-stage radiative cooling system to achieve its desired operating temperatures for the optical
system and FPAs.  The basic approach follows the thermal design of the Planck satellite, which used a 3-stage V-groove
radiator to cool its radiator panels to 140, 90 and 46 K, with the telescope reaching 36 K \citep{Planck_thermal11}. 
However, Planck operated in a halo orbit about the second Earth-Sun Lagrange point (L2), which avoided emission
from the Earth and naturally shaded the V-grooves from direct sunlight by the spacecraft body and solar panel.

Following Planck, SPHEREx has 3 V-groove panels located between the spacecraft and the telescope (see
Figure \ref{fig:spherex}).  Being in low-Earth orbit, the design incorporates 3 conical ``photon shields",
which serve as extensions of the 3 V-groove coolers.  The photon shield cone angles are progressively
increased from inner to outer to allow thermal photons to escape
to space in a few specular reflections.  The nested cones also protect the cold instrument from being illuminated by
the Sun or the Earth.  During observations, the Earth illuminates the inner photon shield, but at an angle where the
light then reflects to cold space without intercepting the instrument.  The telescope and focal planes are supported on
low-conductivity bipods, which are thermally intercepted at each of the 3 V-grooves to reduce conducted power to the
instrument.  The thermally isolated telescope also radiates to space, effectively comprising a fourth radiation stage.
Finally, the MWIR FPA requires the lowest temperature.  It is cooled by a dedicated radiator panel, the fifth
radiation stage, that is isolated from the telescope by bipods.   \citet{Moore25} will present further information
on the design and in-flight performance of the SPHEREx thermal system (also see \S \ref{ssec:thermal_onorbit}).

\subsubsection{Optics}
\begin{figure}
\centering
    \includegraphics[width=\linewidth]{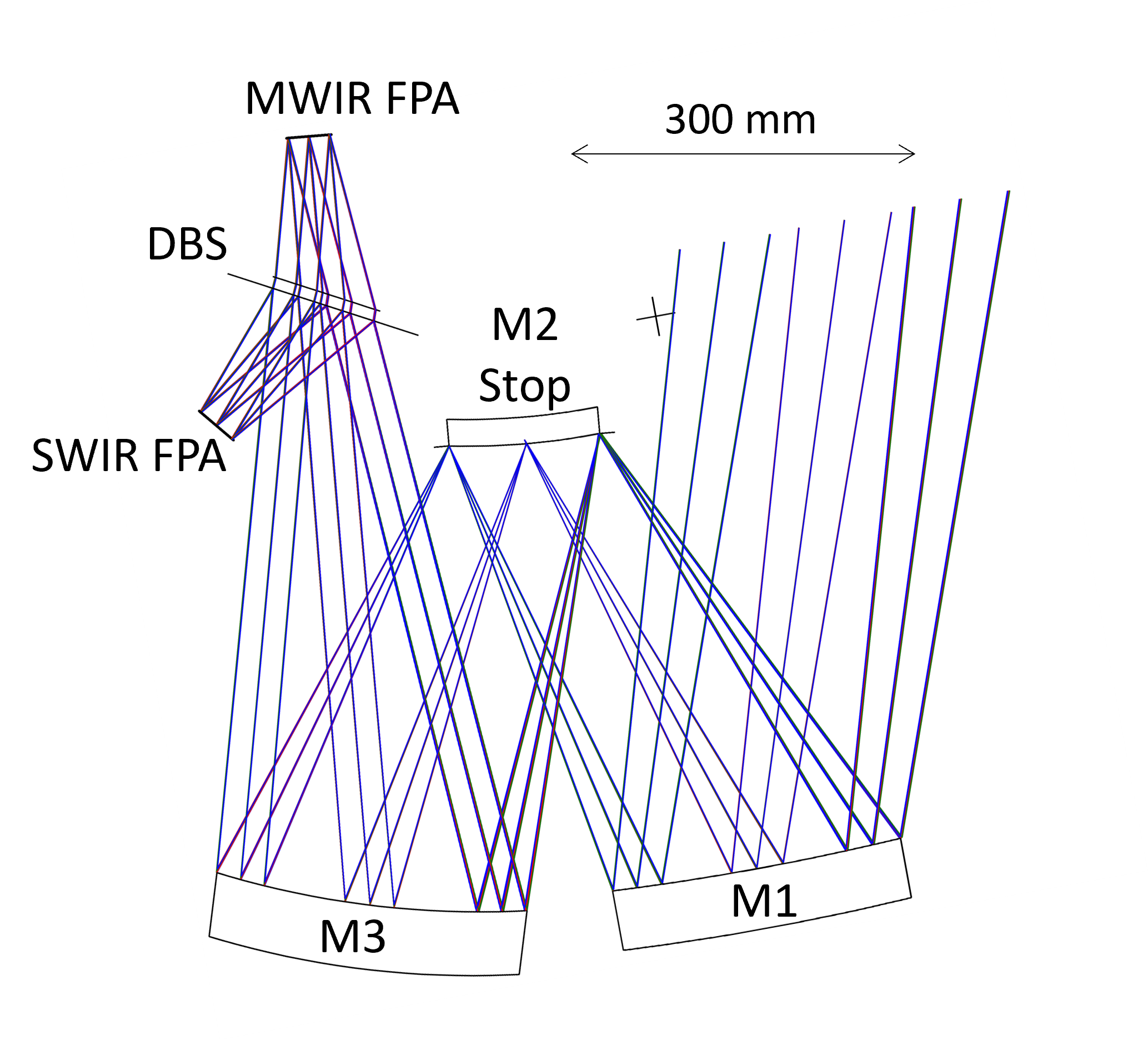}
    \caption{Ray trace of the SPHEREx optics, an off-axis 3-mirror free-form design with a wide
    $3.5^\circ \times 11.5 ^\circ$ field of view.  In this view, the long direction of the field of
    view is out of the page.  The 20.0 cm effective aperture is defined by a pupil stop at the
    secondary (M2), so that the primary (M1) mirror is somewhat physically larger than the effective
    diameter.  The DBS reflects (transmits) infrared light to the SWIR (MWIR) focal plane assemblies
    with an f/3 focus.}
    \label{fig:raytrace}
\end{figure}

\begin{figure}
\hspace{0.3cm}
    \includegraphics[trim={13cm 0 8cm 0.5cm}, clip, width=1.1\linewidth]{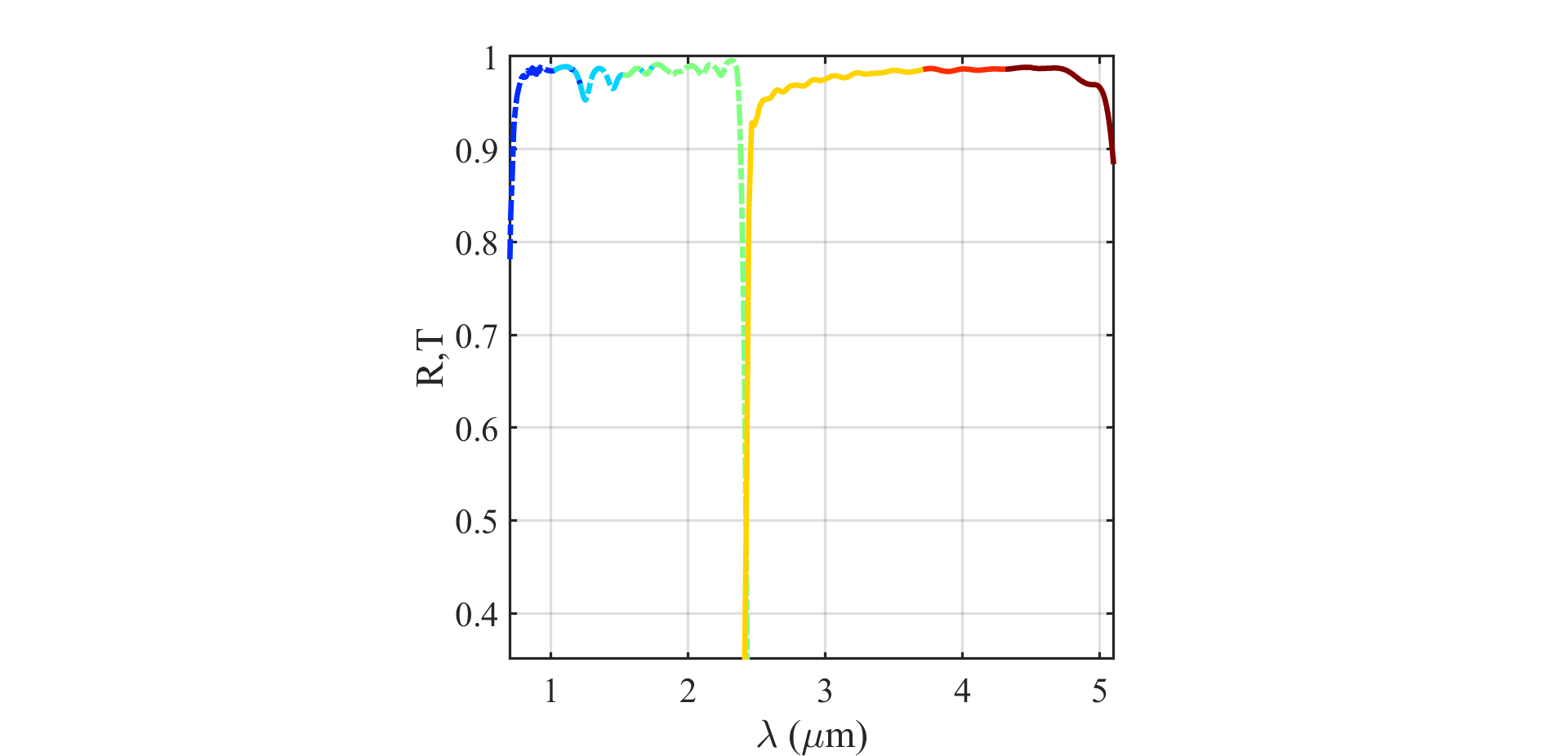}
    \caption{Measured reflection (dashed line) and transmission (solid line) for the DBS, with a crossover wavelength at 2.42 $\mu m$.  The colors indicate the transition between the 6 focal plane bands detailed in Table \ref{table:spectro}.}
    \label{fig:dbs}
\end{figure}

SPHEREx uses an off-axis 3-mirror high-throughput telescope that supports a wide
$3.5^\circ \times 11.5 ^\circ$ telecentric field of view (see Figure \ref{fig:raytrace}).  The optical path
is split by the DBS (see Figure \ref{fig:dbs}) into two focal planes to further increase the light-gathering
power.  The pupil stop at the secondary mirror sets the effective aperture diameter of 20.0 cm.  The illumination
on the primary moves with location on the focal plane, leading to the primary mirror having dimensions of 26.2
$\times$ 35.1 cm.  The optical design minimizes aberrations by using free-form surfaces on all 3 mirrors. 
Assuming ideal alignment, the geometric spot size in the design varies from 3.4 to 11.0 $\mu m$, compared with
18 $\mu m$ pixels, with the largest spots occurring at the corners of the field of view.  The optics have
a moderate level of distortion, up to 0.5 \% over the field of view, which must be accounted in the data processing.

BAE assembled and aligned the telescope, and cryogenically tested the image quality before and after
environmental vibration testing \citep{Frater23}.  The mirrors and housing were all fabricated
from a single billet of Aluminum 6061 alloy in order to minimize misalignment at cryogenic temperature.
The mirrors were fabricated with light-weighted back surfaces and coated with FSG98-protected gold to
reduce reflection loss.  After assembly, the measured spot sizes varied from 3.9 to 10.8 $\mu m$
(SWIR) and 7 to 14.7 $\mu m$ (MWIR) at ambient temperature, and from 4.4 to 17.4 $\mu m$ (SWIR) at
58 K (the cryogenic performance was only measured in reflection off the DBS with optical interferometry).
We attribute the increase, most evident at the extreme corners of bands 3 (see Figure
\ref{fig:Neff}), to alignment errors and mounting stress in the mirrors.  The image quality, assuming
ideal placement of the focal plane surface, is better than \SI{3}{\arcsecond} FWHM over most of the
SWIR focal plane.  In the extreme corners in band 3, the PSF becomes as large as 6\arcsec.
The PSF in the MWIR focal plane increases up to \SI{5.5}{\arcsecond} due to diffraction at longer
wavelengths.

\subsubsection{Spectroscopy}
SPHEREx carries out spectroscopy by placing high-efficiency LVFs over each of
the 6 H2RG focal plane detector arrays.  While this simple approach is unique to astrophysics, it was used
in two NASA planetary missions \citep{Reuter08, Reuter12} to rapidly map the surfaces of solar system objects
in low-resolution infrared spectroscopy.

The LVFs are designed with a logarithmic wavelength progression, such that the distance along the filter
varies as $x(\lambda) = kR \ln(\lambda / \lambda_0)$, where R
is the chosen resolving power (see Table \ref{table:spectro}), $\lambda_0$ is the starting wavelength and
k is the linear distance between spectral channels (11.8 arcminutes), which we hold constant for all
6 of the arrays.  Thus the width of the $i$th spectral channel, between $\lambda_i = \lambda_0 (1+1/R)^i$
and $\lambda_{i+1} = \lambda_0 (1+1/R)^{i+1}$, is $kR \ln(1+1/R) \simeq k$.

\begin{figure*}[!t]
\centering
    \includegraphics[width=18.5cm]{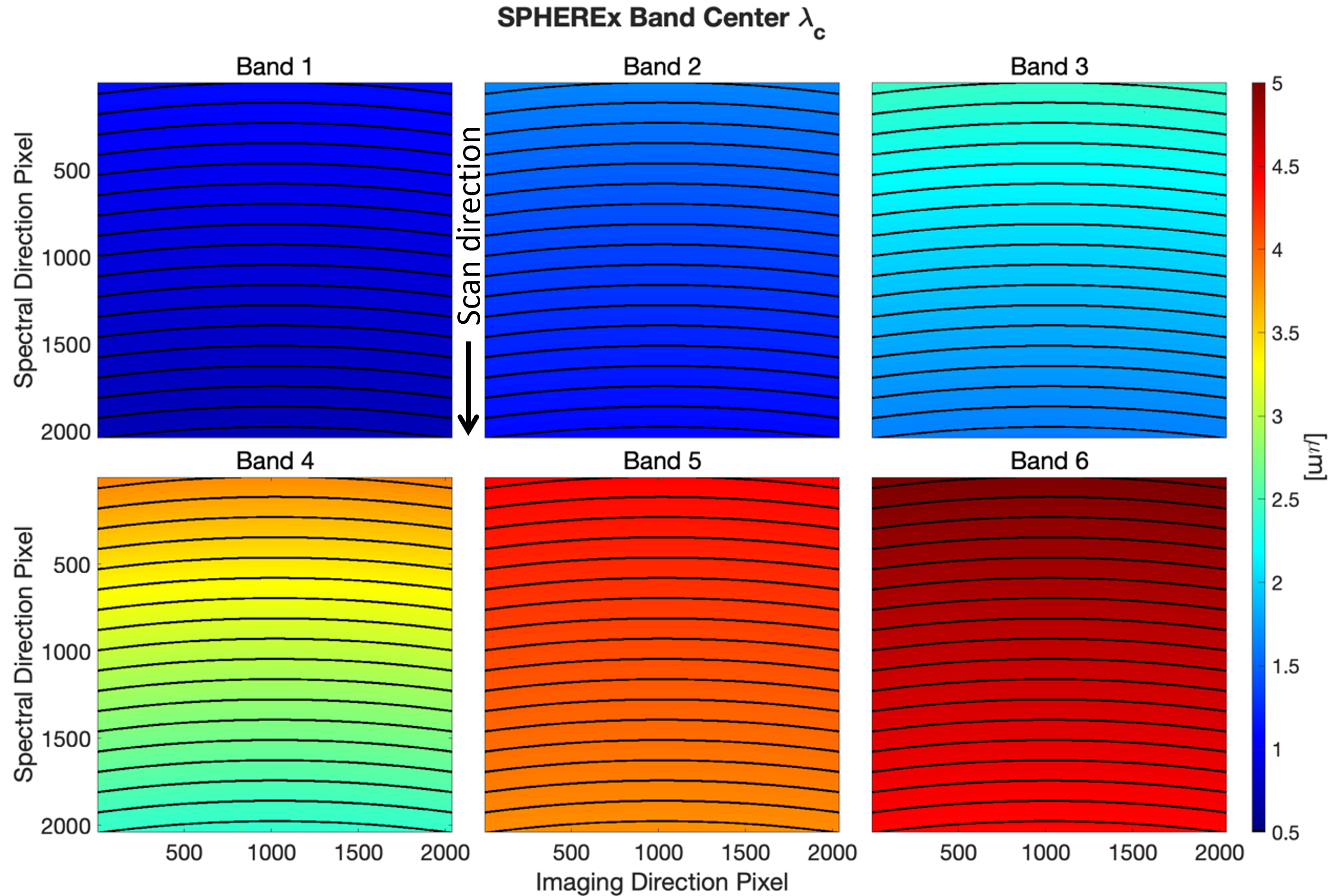}
    \caption{The central wavelength of the spectral response of each pixel measured over the 6 detector arrays in the focal plane.  The central wavelength varies continuously over the focal plane.  We designate with lines our convention for spectral channels.  These regions are defined geometrically, but nearly coincide to where the wavelengths falls between $\lambda_i$ and $\lambda_i (1+1/R)$ for the resolving powers R given in Table \ref{table:spectro}.  The iso-wavelength contours on all 6 LVFs have a matched curvature that comes from the fabrication process.  The survey proceeds by moving the telescope along the vertical direction (indicated by the arrow) between exposures in order to obtain samples in all 102 spectral channels in each survey. Due to the complex survey pattern needed to cover the full sky, the horizontal location of each observation on the field of view will vary object by object depending on its location on the sky.}
    \label{fig:bandc}
\end{figure*}

The spectral response of the instrument was measured at operating temperature in the laboratory using
a specialized cryostat equipped with a cooled integrating sphere and Winston horn arrangement to illuminate
the focal surface uniformly.  The integrating sphere was coupled to a laboratory grating spectrometer
to measure the spectral response of every pixel in the focal plane, providing measurements of the band
width, band center, and out-of-band blocking.  We show the central wavelength measured over the 6
focal plane arrays in Figure \ref{fig:bandc}.  A full description of the measurement apparatus and
results will be presented in \citet{Hui25}.

\subsubsection{Focal Plane Assemblies}
SPHEREx uses two Focal Plane Assemblies (FPAs) to mechanically support and thermally isolate two focal
plane blocks, each with 3 LVF-H2RG pairs.  The LVF is precisely located with a small (50 to 120 $\mu m$)
gap to the detector surface.  The system must have a highly athermal design to maintain
stable focus, to within 0.36 $\mu$m over 6 months, as the telescope temperature changes during
the mission lifetime.  The approach matches the thermal expansion coefficient of the all-aluminum telescope
assembly. The FPA must also keep the detector temperature stable to $< 0.7 ~\mu$K s$^{-1}$ over an exposure to
reduce dark current from thermally-modulated offset voltages.  The FPA thermally isolates the detectors
so that their temperature may be accurately monitored and controlled within the available heat sink power.

\begin{figure}[!t]
\hspace{0.0cm}
    \includegraphics[trim={1cm 0 0cm 0cm}, clip, width=1.1\linewidth]{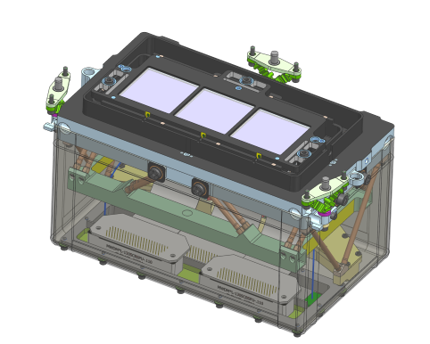}
    \includegraphics[trim={0cm 0 0cm 0cm}, clip, width=1.0\linewidth]{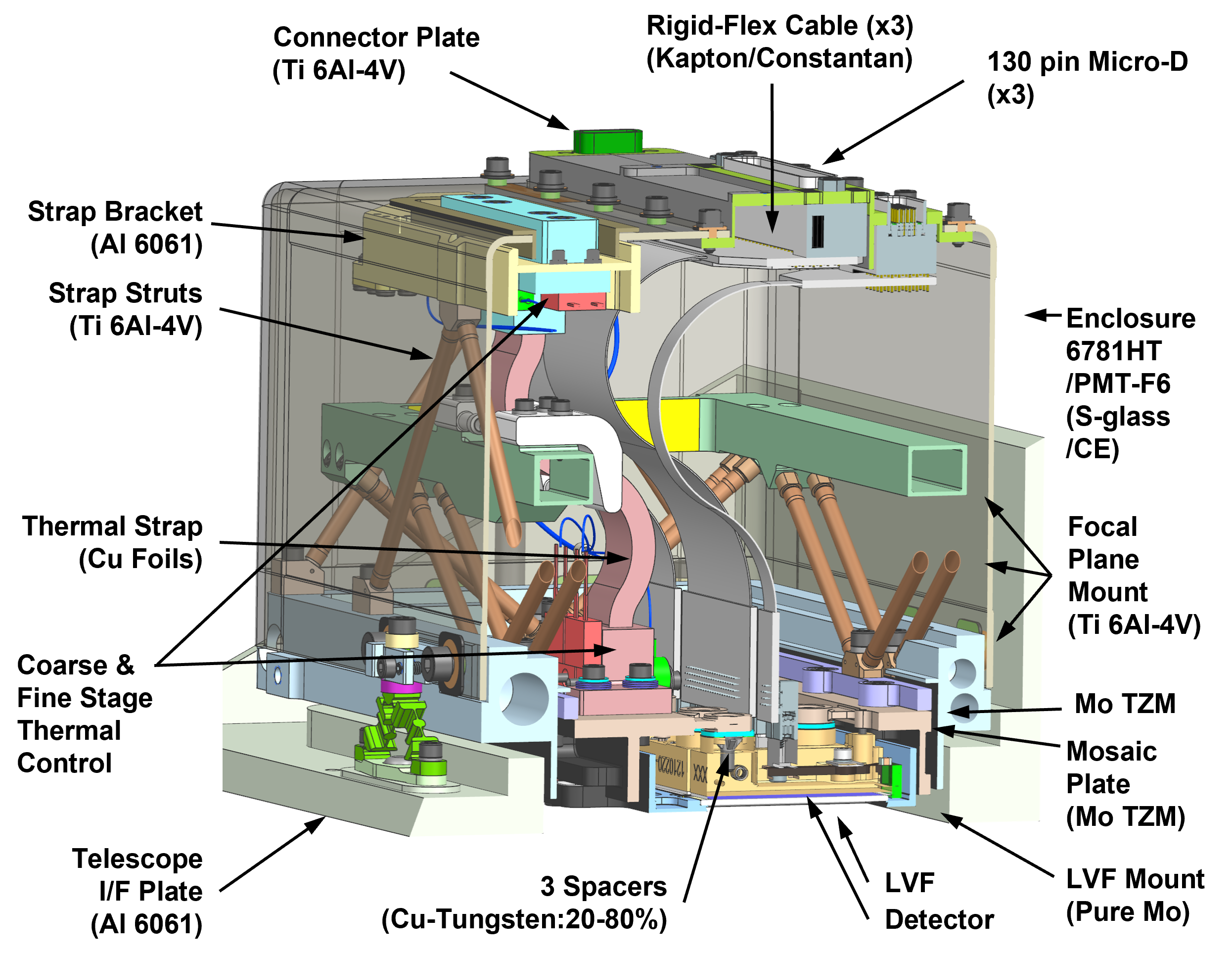}
    \caption{(Top) Overview of the FPA, with the 3 LVF-H2RG pairs pointing up.  The three bimetallic bipod mounting points, where the FPA bolts to the telescope assembly, are also visible at the top surface. (Bottom) Cross-section of the FPA assembly, now inverted with the LVF-H2RG pair at the bottom.  The focal plane block is supported on a reentrant Titanium 6Al-4V bipod assembly with a mid-frame, shown in green.  The thermal strap assembly, shown in pink, is also supported and thermally isolated by a separate set of Ti bipods, providing a bolted interface at the top.  Coarse and fine thermometers and heaters are located at the end fittings of the strap.  A set of kapton cables with constantan electrical traces route the detector clocks and signals to three 130-way connectors at the top.  The enclosure provides micro-meteoroid and ESD protection, and a light-tight environment with labyrinth seals around the thermally-isolated focal plane block and thermal strap.}
    \label{fig:fpa}
\end{figure}

Each FPA (see Figure \ref{fig:fpa}) mounts to the telescope with a 3 mm gap between the mounting points
and the optical focus.  The SWIR focal plane block is thermally connected to the 62 K telescope, but the
MWIR is connected to the MWIR radiator to achieve a lower 45 K operating temperature.
The FPA uses a titanium suspension and Kapton readout cables with constantan electrical traces to provide
the requisite thermal isolation, while the detector temperatures are controlled by a pair of coarse and
fine thermometers and heaters located on the thermal strap.

SPHEREx uses 6 Hawaii-2RG detector arrays \citep{Blank12}, part of a family of detectors with a
long heritage in space-based applications \citep{Beletic24}, housed in the two FPAs.  The SWIR FPA views
the DBS in reflection, while the MWIR FPA views the DBS in transmission.  The SWIR detectors were fabricated
with a cut-off wavelength of 2.65 $\mu$m while the MWIR detectors have a cut-off wavelength of 5.4 $\mu$m.
The properties of the detectors are summarized in Table \ref{table:fpas}.  Note that each H2RG detector
provides a 4 pixel-wide frame of non-photosensitive reference pixels that can be used to assess dark current and
noise properties.

\begin{table}[h]
\hspace{-2 cm}
    \begin{tabular}{c|c|c|c|c|c}
        Band & $Q_{CDS}$ & $\Delta I$ & $I_{dark}$ & $I_{dark,ref}$ & Yield \\
          & $e^- / s$ & $e^- /s$ & $e^- /s$ & $e^- /s$ & \% \\
        \hline
         B1 & 15.62 & 0.057 & -0.060 & -0.018 & 99.89\\
         B2 & 12.12 & 0.048 & -0.050 & -0.017 & 99.88\\
         B3 & 13.67 & 0.053 & -0.058 & -0.022 & 99.82\\
         B4 & 9.16 & 0.051 & 0.023 & 0.044 & 97.07\\
         B5 & 9.27 & 0.049 & 0.023 & 0.049 & 99.15\\
         B6 & 9.37 & 0.048 & 0.019 & 0.043 & 99.09\\                        
    \end{tabular}
    \caption{Properties of the 6 H2RG detectors measured in the laboratory under dark conditions.  $Q_{CDS}$ refers to the median rms charge on a pixel after differencing two consecutive reads, $\Delta I$ is the median error in the current in a 112 s integration, $I_{dark}$ is the median dark current without reference pixel subtraction, $I_{dark,ref}$ is the dark current after subtracting the median of the reference pixels in each readout channel, and the yield is the fraction of pixels that were designated as usable for science observations.  Most of the measured dark current comes from a transient response to the reset where the response is somewhat different between reference and optical pixels.  The dark current also arises from operating the detectors in 32-channel output mode with a rapid 1.5 s read interval, which produces a low level of multiplexer glow.  These values include any contributions to dark current and noise from the warm readout electronics.}
    \label{table:fpas}
\end{table}

\subsubsection{Readout Electronics}
\label{sssec:electronics}
SPHEREx warm electronics (see \cite{Heaton23} for a detailed description) control and read the detector
arrays, digitize the data, perform sample-up-the-ramp (SUR; \cite{Garnett93}) fitting to extract the detector
current, compress the resulting data, and pass the output to the spacecraft computer for telemetry.
The electronics also reads out instrument thermometry and controls the SWIR and MWIR temperatures.
The readout electronics employ 6 identical readout boards, which provide bias and timing signals
for each of the 6 detector arrays.  The readout boards are operated by a central electronics board
that communicates with the spacecraft computer.  The bias voltages for the H2RGs are generated in
an oven-controlled region of the readout board to minimize thermally-induced drifts.

Each H2RG detector is read out by 4 custom Video8 application-specific integrated circuit (ASIC) amplifier chips,
operating 32 readout channels at 100 kHz sampling.  The Video8 was designed for low noise
and high $1/f$ stability, commensurate with the diffuse mapping requirements of the galaxy formation theme.
Each Video8 chip has 8 fully differential readout channels, designed to minimize EMI/EMC and microphonic
susceptibility of the readout cables between the warm electronics and the cryogenic detectors.  The
ASIC has two banks of 4 readout channels, where each bank interfaces with a 20-bit off-chip ADC.  A Video8
readout channel consists of a preamplifier and a pair of integrators.  The integrators filter the signal
for optimum noise performance, and then sample and hold the signal for the ADC.  The integrators work in
an interleaved fashion, with one side carrying out the integration while the other holds for the ADC.  The
design also features analog switches at the preamplifier input, where the switches can be closed to a
reference voltage to monitor the offset of the amplifier.  These reference samples are incorporated into
the data stream as ``phantom pixels'', and are used to monitor and precisely subtract the leakage current of the
preamplifier (typically several $e^-/s$).  Caltech and JPL qualified the Video8 ASIC for operation in space,
including radiation tests for latch-up and total dose.

SPHEREx uses the SUR algorithm to reduce the data volume to fit within
the available telemetry bandwidth.  SUR provides a best-fit slope to measure the photocurrent using a
constant sampling cadence every 1.5 s.  The regular sampling provides additional flexibility, enabling the
detection of transients and saturation on board.  Our SUR algorithm \citep{Zemcov16} identifies saturation
on bright sources, and for sources that do not saturate in a few reads, returns a best-fit slope over a
restricted time period to give a larger dynamic range.  The algorithm also searches for voltage jumps caused
by energetic particles, to a chosen threshold, and returns a best-fit slope from data prior to the event.
The algorithm returns flags identifying these special cases, and also a flag indicating if the flagged voltage
jump occurred in the first half or second half of the exposure.

SPHEREx uses a ``row-chopping'' method to mitigate large-scale spatial noise, as required by the galaxy formation
theme.  Row-chopping reads rows non-sequentially, $i.e.$ reading a row, skipping 32 rows, and reading the next
row.  The effect is to modulate common-mode $1/f$ noise, dominated by the amplification chain in the the array
read out integrated circuit (ROIC), to high spatial frequencies.  The modulated $1/f$ noise is small at these
frequencies compared with white noise.

Row-chopping is very effective in removing the effect of $1/f$ noise within a readout channel.  However, the
common-mode $1/f$ leaves a small drift that translates to a different offset current in each readout channel.
We can average the 8 rows of photo-insensitive reference pixels available in the H2RG detector to measure
and subtract the offsets in each channel.  Laboratory tests in a dark test cryostat \citep{Nguyen25} using
this method showed the low spatial frequency noise was below the expected level of photon noise in bands
1 - 4 for spatial wavenumber $k > 0.03 ~{\rm pix}^{-1}$ (or multipole $\ell > 500$ for our 6\farcs2
pixel scale).  Note the photon noise level in Figure 9 of \cite{Nguyen25} is from an early estimate of
optical efficiency, and should be increased by a factor of $\sim$4 (in $e^2 ~s^{-2}$ units) to match the
in-flight photocurrent.  The tests also showed that subtracting the median of the optical pixels within
a  readout channel further improves the $1/f$ performance, as the channel offsets can be measured more accurately
(even with photon noise) than by averaging the comparatively small number of reference pixels.

\subsection{Spacecraft}
The SPHEREx spacecraft (see Figure \ref{fig:spherex}) is based on a single-string architecture using
high-reliability components.  The basic function of the spacecraft is to reliably return spectral
imaging data from the instrument by executing the survey plan with a succession of agile slews to accurate
and stable pointings, while maintaining the avoidance criteria to the Earth and Sun.  The spacecraft is
based on a modular configurable platform developed by BAE Systems.

The hexagonal bus structure supports the payload and houses the spacecraft components on its interior
and exterior sides.  Components mounted on the exterior include the instrument readout electronics,
two star tracker optical heads, the single solar panel array, and the inertial reference unit and
magnetometer.  Given the dimensions of the Falcon 9 launch vehicle fairing, the project chose a design
with fixed structures for the photon shields and solar panel to eliminate the risk of mechanical
deployments.

The attitude determination and control system measures attitude and rates by combining a two-headed star
tracker to provide an absolute celestial reference, with an inertial reference unit based on a hemispherical
resonator gyro.  An array of sun sensors, one magnetometer and an on-board GPS receiver provide reliable
coarse attitude determination, celestial positioning, and timing knowledge.  The system uses a set of 4 reaction
wheels to execute maneuvers for slews and pointing stabilization.  The wheels are mounted in a tetrahedral
configuration oriented to maximize slew performance along the axes most commonly used in survey operations.
The configuration allows the vehicle to retain three-axis control, although with reduced slew performance,
on 3 wheels.  The spacecraft intermittently unloads momentum during slews and communication
maneuvers through the Earth's magnetic field via 3 magnetic torque rods. SPHEREx does not require orbit
maintenance and therefore has no propulsion system.

The spacecraft receives compressed data from the instrument that it stores on a 128 GB flight data recorder
with more than 6 operating days of science storage capacity.  The spacecraft communicates via S-band for command
uplink and health-and-status downlink, and Ka-band for high-rate science data downlink.  The S-band
transceiver provides 2 Mbps in pointed downlink via a nadir-pointed antenna, with omnidirectional
coupled antennas for 32 kbps downlink and 2 kbps command uplink.  The Ka-band system transmits at
600 Mbps through two medium-gain antennas located at the bottom of the spacecraft.  The boresights of
the two antennas are offset to give sufficient views to ground stations near the Earth's poles while
maintaining the Earth and Sun avoidance criteria.  Only one of the antennas transmits at a time through
a radio frequency switch.

The spacecraft avionics unit houses both general and mission-specific cards.  The RAD750 single-board
computer within the avionics unit hosts the flight software, which operates the spacecraft subsystems
using a table-driven software architecture.

\subsection{Mission Operations}
Mission operations are conducted at JPL from the Earth Orbiter Mission Operations Center using ground systems
adapted from the WISE/NEOWISE missions.  The Survey Planning Software \citep{bryan25}, developed by
Arizona State University, is run for the next planning period using the latest orbit prediction to provide
an accurate forecast of Earth and Moon avoidance constraints.  The team generates command sequences twice
a week from the survey plan, and checks constraints independently.  During a typical day, the spacecraft transmits
science data through several $\sim$5 minute passes.  The spacecraft points during these maneuvers to maintain the 
telemetry link while adhering to the avoidance constraints.  No imaging data is taken during these downlinks.
The team issues commands and receives telemetry via the S-band link.   The mission operations team typically
downlinks 20 GB of science data each day at 600 Mbps via the Ka-band link.  All communications are routed through
NASA's NSN ground stations, including those at Fairbanks in Alaska, Punta Arenas in Chile, Svalbard in
Norway and the Troll Satellite Station in Antarctica.

\subsection{Data Processing and Archiving}
\label{ssec:pipeline}
\begin{figure*}
\hspace{-0.1cm}
    \includegraphics[trim={0cm 0cm 0cm 0cm}, clip, width=1.0\linewidth]{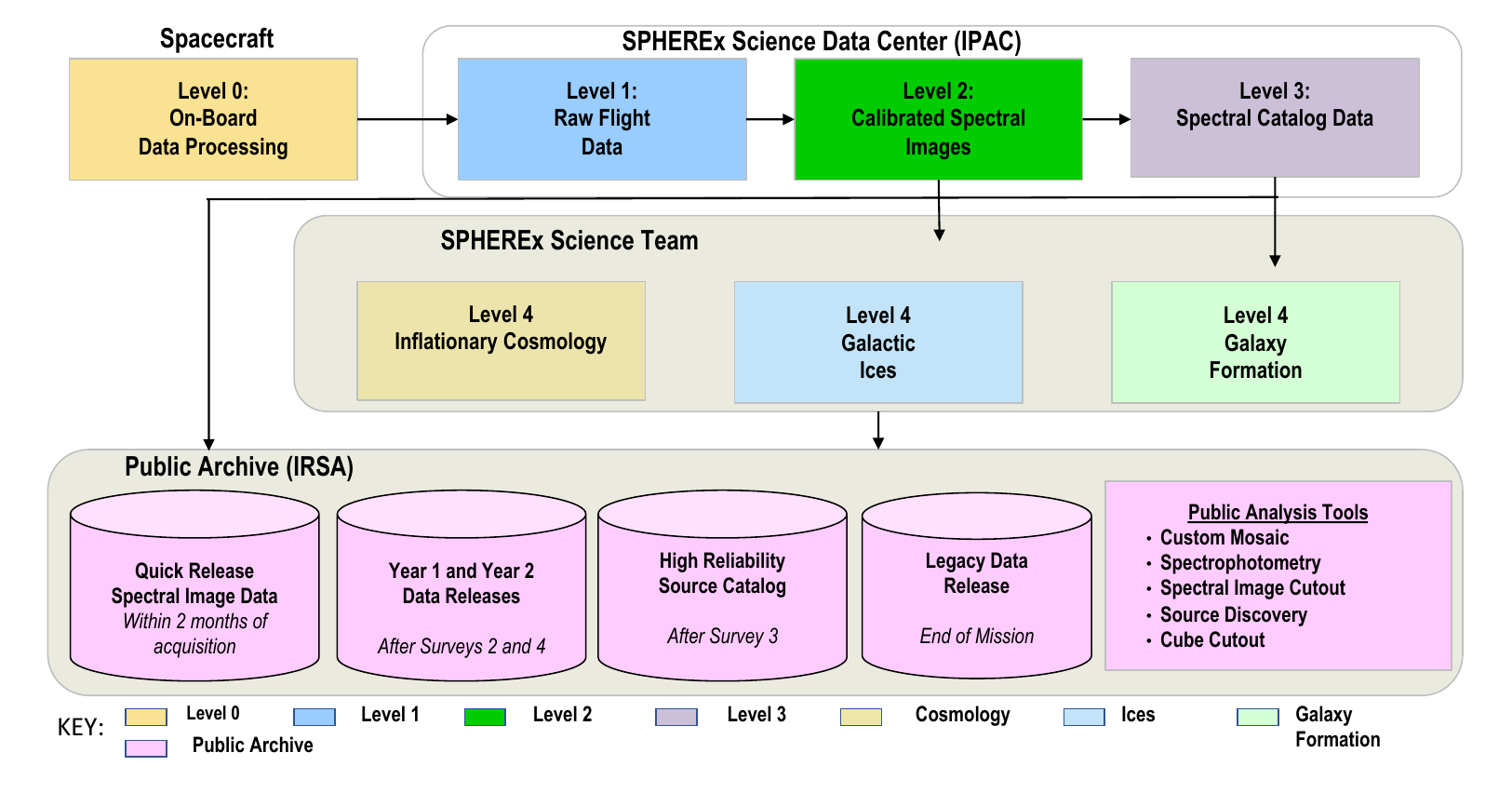}
    \caption{Schematic of the SPHEREx data processing pipelines, data products and tools.}
    \label{fig:pipeline}
\end{figure*}

The SPHEREx data processing flow includes four levels of data processing.  Level 1 to 3 are performed by the
SPHEREx Science Data Center (SSDC) at IPAC and Level 4 is performed by the SPHEREx Science Team.   Importantly,
the SSDC pipeline does not perform blind source detection and extraction. Instead, it measures the flux of sources
listed in a reference catalog \citep{Yang2025} that is generated by the SPHEREx Science Team which includes
calibration sources (both
photometric and astrometric), science targets for the cosmology and interstellar ices themes, and any nearby
sources that need to be included for photometric separation. Figure \ref{fig:pipeline} shows the high-level
organization of the pipeline levels, data products and tools. Additional details for the Level 1 to 3 pipeline
can be found in \citet{Akeson25b} and the SPHEREx Explanatory Supplement \citep{akeson25}.

\subsubsection{Level 0 and 1 Processing}
The Level 1 pipeline takes Ka-band and S-band inputs from data ingestion and produces
engineering unit images in a standard format and with the necessary header information for later calibration
operations.  Key functional steps in Level 1 include:
\begin{itemize}
    \item Decompressing the individual packets of raw data and re-ordering according to the ``row-chopping'' readout scheme (see \S \ref{sssec:electronics}).
    \item Merging S-band health-and-status data and Ka-band science data. 
    \item Estimating initial pointing information based on the survey plan and geometric offsets.
    \item Validating the on-board SUR algorithm (see \S \ref{sssec:electronics}) by comparing the full readout sequence for every pixel of a sub-section of the detector to the best-fit slope downlinked to the ground.
    \item Converting raw pixel data from analog digital units (ADU) to electrons per second.
    \item Applying corrections for amplifier chain drifts using reference pixels and interleaved phantom pixels. 
    \item Converting to FITS and performing basic checks on data integrity, including image statistics.
\end{itemize}

\subsubsection{Level 2 Processing}
Level 2 processing performs astrometric and photometric calibration using a combination of ancillary
catalogs and internally generated calibration products from the pipeline, and sets other calibrations
and flags. The outputs of Level 2 are calibrated spectral image files, which after data quality checks, are
rapidly released as data products through the NASA/IPAC Infrared Science Archive (IRSA).  The Level 2
data files remove reference and phantom pixels.  Key functional steps in Level 2 include:
\begin{itemize}
    \item Correcting for detector non-linearity, which arises from declining gain as charge accumulates within each H2RG pixel \citep{Zengilowski20}.  
    \item Creating a per-pixel variance map from the measured flux. 
    \item Applying a parameterized analytic model to estimate the level of persistent flux in each pixel based on the cumulative flux exposure of that pixel.  
    \item Generating a per-pixel flag map including transients, non-functional pixels, outlier pixels, persistence and optical ghosts.
    \item Applying the dark current, flat-field and absolute gain calibrations to the image, and converting data values to units of MJy/sr.
    \item Calculating an astrometric solution to determine the translation and rotation of the field per image using known isolated stars from the Gaia catalog as references.  
    \item Creating a source mask for sources in the reference source catalog created by the SPHEREx Science Team.
    \item Detecting and flagging transient events that were not flagged by onboard procedures, using a sigma-clipping algorithm applied to median-filtered images.  
    \item Providing an estimate of the diffuse Zodiacal light \citep{Crill2025} based on the spatial location of the image.  The empirical model is based on \cite{Kelsall98}, using spectral measurements from \cite{Tsumura10, Tsumura13}. 
    \item Creating the spectral image as a multi-extension FITS (MEF) file providing the image in MJy/sr, per pixel data flags, estimated noise in (MJy/sr)$^2$, estimated Zodiacal background from the modified Kelsall model (which is provided for reference but not subtracted from the image), the point spread function on an 11 x 11 pixel grid across the detector, and a binary lookup table with central wavelengths and bandwidths.
    \item Performing a series of automated tests on the output from each module and on the final spectral image FITS file.  The image must pass the data quality test, be outside the SAA, have a transient count less than 435,000 pixels and have a fine astrometry flag value of 0 in order to be released through IRSA.
\end{itemize}

In addition, the Level 2 pipeline derives the absolute gain matrix by combining both the flat-field and the dark
current matrices with measured calibration factors.  As described in \S \ref{sssec:photometric},
we derive the flat-field and dark current matrices using hundreds of input images obtained over varying Zodiacal
light levels.  We use observations of pre-selected primary calibrator stars to derive the calibration
factors by comparing the measured and model flux as a function of wavelength (see \cite{akeson25} for
the list).  We apply the calibration per pixel using the absolute gain matrix and the dark current
calibration products.  The pipeline determines the sub-pixel PSF in a
grid over each band by stacking and deconvolving the measured response for a large number of isolated
and high-SNR stars.   The resulting product gives an effective PSF in each region that includes both the
intrinsic optical response and the average blurring due to pointing jitter.   The original design separated the
optical PSF and the per-observation jitter, but the measured jitter is so low (see Figure \ref{fig:settle})
that we only use the average PSF per region. IRSA provides the absolute gain matrix, the dark current matrix
and the derived PSFs as calibration products. The PSFs are also available as an extension within each
Level 2 calibrated file.

\subsubsection{Level 3 Processing}
The Level 3 pipeline applies an optimal photometry algorithm on the calibrated spectral images produced by
Level 2 to measure the flux at the positions of pre-selected sources contained in the reference catalog built by the SPHEREx Science Team.  The PSF photometry method is based
on the community Tractor (\url{thetractor.org}) package that includes background subtraction and error
covariance from overlapping PSFs. The All-Sky Spectral
Catalog assembles these photometric outputs, both as measured from individual spectral images and binned
to a grid of 102 wavelengths.

\subsubsection{Level 4 Processing}
The 3 core science themes carry out their analyses starting with Level 2 and 3 products.  The cosmology theme
starts from Level 3 optimal photometry on known galaxy positions to produce the galaxy redshift catalog.  The
galaxy formation theme takes Level 2 calibrated spectral images and mosaics these into the deep-field
maps.  The ices theme starts from Level 3 optimal photometry on a target list of background and embedded stars to
produce ice absorption spectra.  The legacy catalogs also run mainly on Level 3 photometry, with additional
considerations for moving solar system objects and source crowding in galaxy clusters.

\subsubsection{Data Products and Tools}
\label{sec:dataproducts}
IRSA provides public access to the SPHEREx science data products given in Table \ref{table:dataprod}.
There are two planned reprocessing periods, after Year 1 and after Year 2.  All data will be reprocessed
with the most recent pipeline version and calibrations.  Data products from Year 1 and Year 2 will be
available within 6 months after the end of data collection.  The all-sky data cubes are spatially and
spectrally interpolated to provide all-sky maps at 102 wavelengths.  We will release a subset of the
all-sky catalog as the High Reliability Source Catalog (HRSC), with sources selected to be consistent
between surveys that have a minimum SNR in at least a subset of wavelength bands. The photometry in the
HRSC will be made available both at measured wavelengths and binned to a grid of 102 wavelengths.

\begin{table*}[h]
    \centering
    \begin{tabular}{p{2.3in}|p{1.85in}|p{2.35in}}
        Product & Schedule & Notes \\
        \hline \hline
	Quick Release Calibrated Spectral Images & Within two months of acquisition & Updated weekly at IRSA \\ \hline
	Calibration Products & As needed & Calibrations used with quick release calibrated images \\ \hline
	Reprocessed Calibrated Spectral Images  & Year 1 and Year 2 data releases & Cumulative re-processing of spectral image data \\ \hline
	All-sky Data Cubes  & Year 1 and Year 2 data releases \\ \hline
	High Reliability Source Catalog & 8 months after end of survey 3 \\ \hline
	Deep Field Mosaics & May 2028 & Galaxy Formation science theme \\ \hline
	Stellar Type/Ice Column Density Catalog & May 2028 & Interstellar Ices science theme \\ \hline
	Redshift Catalog & May 2028 & Cosmic Inflation science theme \\ \hline
	Legacy Catalogs: Stellar/Brown Dwarfs,  Galaxy Clusters, Solar System & May 2028 & Science Team Legacy Catalogs \\ \hline
    \end{tabular}
    \caption{Availability of SPHEREx Science Data Products.  Note that Year 1 observations end in May 2026, and Year 2 observations end in May 2027. The Year 1 and Year 2 data releases are within six months following these dates.
    \label{table:dataprod}}
\end{table*}

In addition to the general capabilities of data search, retrieval and visualization, IRSA will host
several SPHEREx-specific tools to facilitate community usage of the data (see the SPHEREx pages at
IRSA for more details e.g. \url{https://irsa.ipac.caltech.edu/data/SPHEREx/docs/overview_qr.html}).
Table \ref{table:IRSAtools} briefly describes the tools, along with their planned availability.

\begin{table*}[h]
\centering
\begin{tabular}{p{1.35in}|p{0.9in}|p{4.25in}}
Tool & Availability & Description \\
\hline \hline
Spectrophotometry & August 2025 &   Measure spectra at user-supplied positions, using the same PSF photometry method as the Level 3 pipeline \\ \hline
Spectral Image Cutout & October 2025 &  Select sections of spectral images based on user criteria (spatial area, band, time) \\ \hline
Source Discovery &  October 2025 & Identify significant signal in user-supplied spatial region (with user constraints), without priors on the position \\ \hline
Custom Mosaic & February 2026 &  Create single-wavelength images from spectral images using user supplied criteria, including synthetic bands \\ \hline
Spectral Cube Cutout &  November 2026 & Extract subsets from the All-sky Spectral Cube, returning FITS cubes with a HEALPix projection, and optionally interpolate for synthetic bands \\ \hline
\end{tabular}
\caption{Availability of SPHEREx-specific analysis and visualization tools at IRSA}
\label{table:IRSAtools}
\end{table*}

\section{On-Orbit Performance}
\label{sec:onorbit}

SPHEREx was co-launched with the 4 NASA PUNCH (Polarimeter to Unify the Corona and Heliosphere) satellites
from Vandenberg Space Force Base on 11 March 2025 8:10 pm PDT (12 March 2025 03:10 UTC) on a Falcon 9
rocket.  The satellite went through a 50-day in-orbit checkout (IOC) period prior to starting science
observations on 1 May 2025.  The IOC was used to first decontaminate the instrument, eject the telescope
cover, cool to operating temperature, and carry out a series of characterization measurements of the
spacecraft and instrument.

\subsection{Decontamination Procedure}
Following launch, the satellite went through a decontamination procedure, whereby the spacecraft maintained
a fixed inertial pointing to drive off volatiles, especially water, from the optics, filters and detectors. 
The MWIR focal plane, designed to measure water ice in the interstellar medium, is particularly susceptible to
any residual water ice in the instrument, where 20 nm of accumulated ice would produce a $\sim$6 \% absorption depth
in the 2.9 - 3.5 $\mu m$ ice feature.  The inertial configuration causes the telescope to alternately view
the Earth and space, cycling the temperature of the telescope and the MWIR focal plane, as shown in Figure 
\ref{fig:cooldown}.

The telescope cover was ejected on 18 March 2025, pointing the spacecraft in the
anti-ram direction to avoid any possibility of a later collision.  After lid ejection, the spacecraft
was pointed to local zenith, enabling radiative cooling to cool the instrument.  For the early phase of
the cooldown, we applied heater power to the telescope optics and MWIR focal plane to keep their temperatures
slightly above that of the telescope housing, in order to prevent ice from accumulating on these critical
surfaces.  The heaters were turned off once the telescope reached 150 K.  We note that the dichroic beam
splitter is relatively thermally isolated, and naturally lags behind the telescope without the application
of heater power.

To validate the decontamination procedure, we compiled observational data of the diffuse Zodiacal sky during
IOC.  Assuming the instrument calibration measured in the laboratory prior to flight, we found no evidence
for the characteristic ice absorption feature at 2.9 - 3.5 $\mu m$ at the 1 \% level.

\subsection{Thermal Performance}
\label{ssec:thermal_onorbit}
\begin{figure}[th]
\centering
    \includegraphics[width=\linewidth]{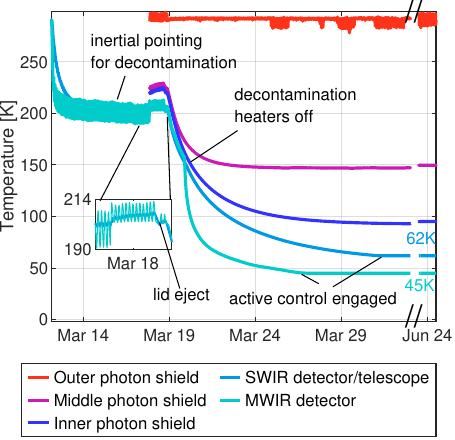}
    \caption{Instrument cooldown after launch, shown in selected instrument temperatures.  For the first 6 days the spacecraft remained inertially pointed, cycling the instrument in temperature as the telescope alternately viewed the warm earth and cold space each orbit.  After this decontamination period the telescope cover was ejected and the spacecraft acquired zenith to maximize radiative cooling.  The spacecraft operated heaters on the optics and the MWIR focal plane to keep these temperatures above the telescope housing.  Once the telescope temperature reached 150 K, the decontamination heaters were switched off.  Finally we engaged thermal regulation heaters to stabilize the SWIR detectors at 62 K and the MWIR detectors at 45 K.}
    \label{fig:cooldown}
\end{figure}

Once the decontamination heaters were turned off, the instrument radiatively cooled to operating temperature
as shown in Figure \ref{fig:cooldown}.  The multi-stage thermal design achieved the expected temperatures of
290 K, 150 K, and 90 K on the 3 photon shields.  Once the MWIR focal plane reached 45 K, we engaged a control
heater to stabilize its temperature.  The telescope cooled more slowly, as expected.  Once it reached 62 K, we
engaged a second control heater to stabilize the SWIR focal plane temperature.  After cooling stabilized,
we observed that the control heaters maintained these temperatures with 47 and 234 mW dissipation for the MWIR
and SWIR focal planes, respectively.  Without heater control, our thermal model predicts that the telescope
and MWIR focal plane would reach 56.5 - 58.4 K and 38.6 - 39.7 K respectively.

Over the first 2 months of scientific operation, the MWIR and SWIR focal plane temperatures
have remained stable to within $\pm$100 $\mu$K, judging from thermometers not used in the control loop. 
During this time we found that the rate of change of the MWIR focal plane temperature was less than 40 nK/s.
Given the susceptibility we measured to temperature in the laboratory, ~150 ${\rm e}^- /{\rm K}$ for both
the SWIR and MWIR detectors, thermal variations induce a negligible $\sim$6 $\mu {\rm e}^-/{\rm s}$
detector dark current.

\subsection{Slew and Pointing Performance}
\begin{figure}[th]
\centering
    \includegraphics[width=\linewidth]{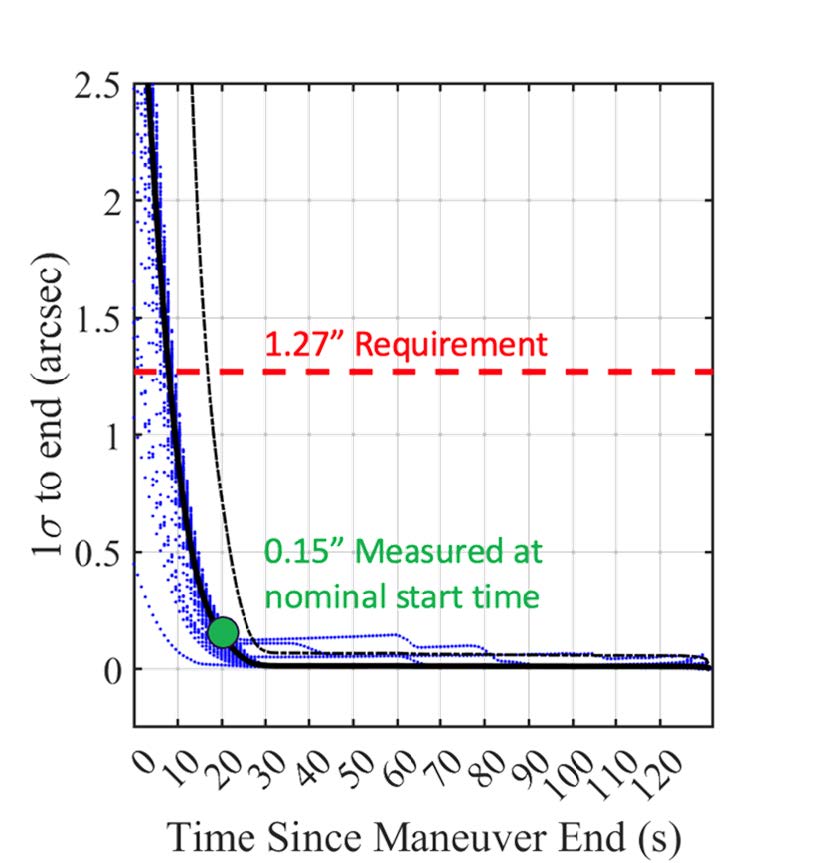}
    \caption{Measured pointing stability following a large slew.  The rms variation is given during the 
    observation as a function of the settling time.  With the chosen settling time of 19.7 s, the pointing
    stability of 0\farcs15 easily meets our requirement of 1\farcs27 that was set to hold pointing smear to an acceptable level.  We chose not to reduce the settling time further however, due to the variability in the settling time evident in the family of trial curves.}
    \label{fig:settle}
\end{figure}

The spacecraft demonstrated excellent slew agility and pointing stability on orbit.  After measuring the
performance of slews, the survey allocates $19.7 + 1.088 \theta_{slew}$(deg) seconds for large slews, 
where a typical 35 deg slew requires 38.1 s for the slew motion plus 19.7 s to settle.  The observed
pointing stability of \SI{0.15}{\arcsecond} rms (see Figure \ref{fig:settle}), with the
same settle time, produces only a small, sub-percent broadening of the optical PSF in a 116.9 s integration.

\subsection{Point Spread Function}
\begin{figure*}
\hspace{-0.0cm}
    \includegraphics[trim={0cm 0cm 0cm 0cm}, clip, width=1.0\linewidth]{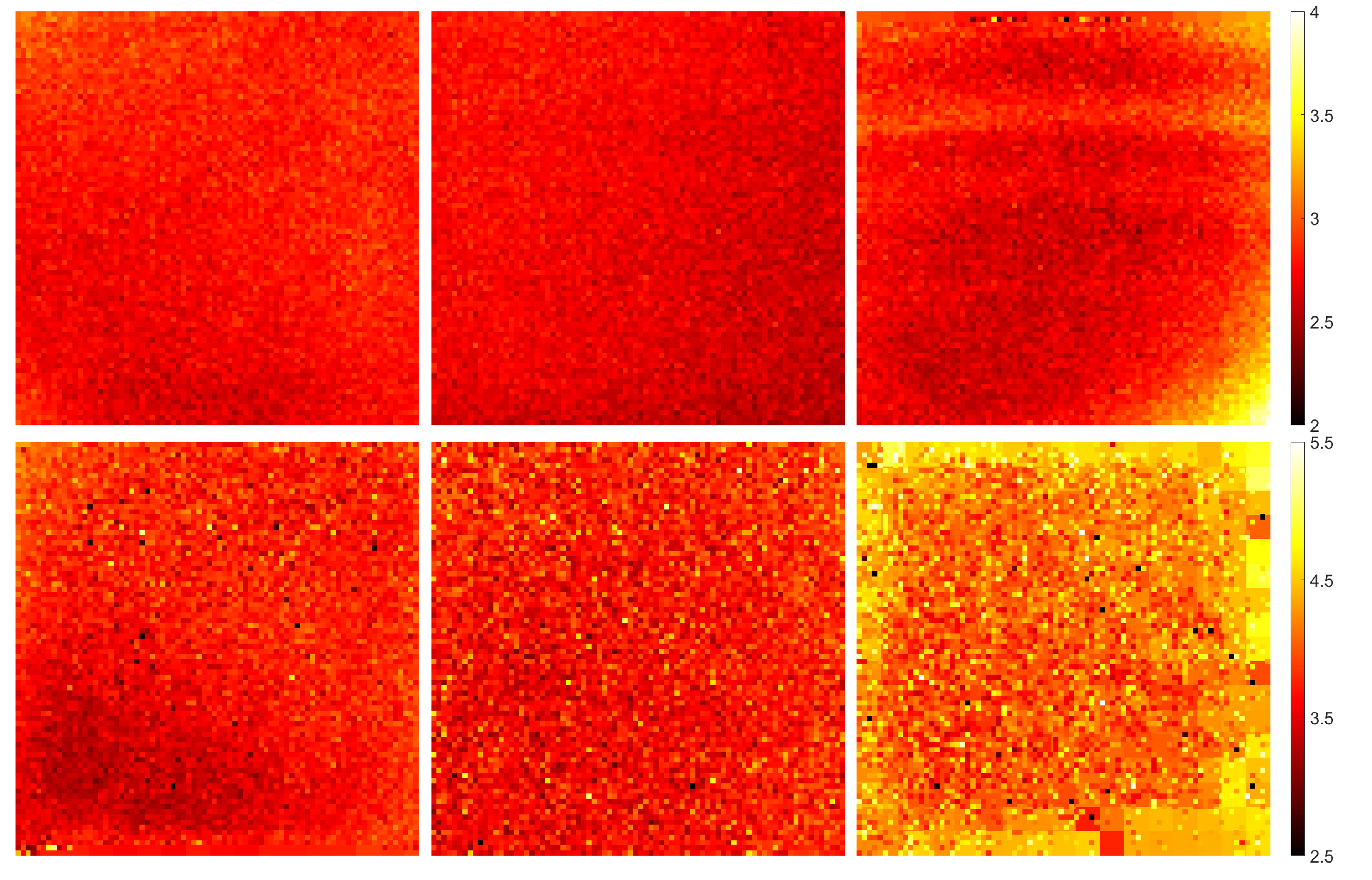}
    \caption{Distribution of the median $N_{eff}$ over the focal planes, in units of number of effective pixels, going from bands 1-3 (left to right on top) and bands 4-6 (left to right on bottom).  Note the color scale for the top 3 images is different from the bottom 3 images. This map was produced from a collection of 1.2 million stars giving $150 < i < 300$ $e^-/s$, and calculating the quantity $\Sigma (1/p_i^2)$ for each star, where $p_i$ is the fraction of total flux in each pixel.  The median of $N_{eff}$ is then shown in each sub-region of the focal plane.  The amplitude of $N_{eff}$ gradually increases with wavelength due to diffraction.  The increase at the corners of the field of view, especially notable in band 3, is due to a moderate level of optical aberration.  The horizontal banding in band 3 may be due to variations in the DBS index of refraction.  Some averaging is evident in band 6 due to a lower density of high-SNR stars in this band.}
    \label{fig:Neff}
\end{figure*}

We measure the response to stars to characterize the in-flight point spread function using two methods.  The
simplest technique is to characterize the effective number of pixels $N_{eff}$ in a star image by calculating
the relative signal in each pixel using $N_{eff} = \Sigma (1/p_i^2)$.  An advantage of this method is that
it is insensitive to astrometry errors.  The measured map of the median $N_{eff}$ shown in Figure
\ref{fig:Neff} agrees well with pre-flight measurements of the telescope image quality.

We also estimate the underlying PSF by stacking on stellar positions to sub-pixel precision,
and then deconvolving the pixel kernel function to obtain an estimate of the PSF in small regions
of the focal plane.  We can then reobserve this estimated PSF in randomized sub-pixel positions to
obtain a map of the median $N_{eff}$.  We find the estimated PSF gives a median $N_{eff}$ map that
closely resembles Figure \ref{fig:Neff} but with $\sim$22 \% larger values, suggesting that the
sub-pixel stacking process introduces a modest level of additional broadening.  We expect to see
closer agreement as our image processing algorithms and astrometry improve. In the future, the
sub-pixel stacking approach may be expanded to quantify further effects, such as crosstalk and
sub-pixel response functions.

\subsection{Gain and Dark Current}
We produced preliminary in-flight estimates of the flat-field gain and dark current using an
ensemble of images taken over varying Zodiacal brightness levels (see \S \ref{sssec:photometric}).
Both estimates closely resemble measurements obtained during laboratory testing of the instrument and
focal planes.  As we refine the algorithms and include more data, \cite{akeson25} will contain the
most recent estimates.

\subsection{Noise Stability}
\begin{figure*}
\hspace{0.2cm}
    \includegraphics[trim={0cm 0cm 0cm 0cm}, clip, width=0.95
    \linewidth]{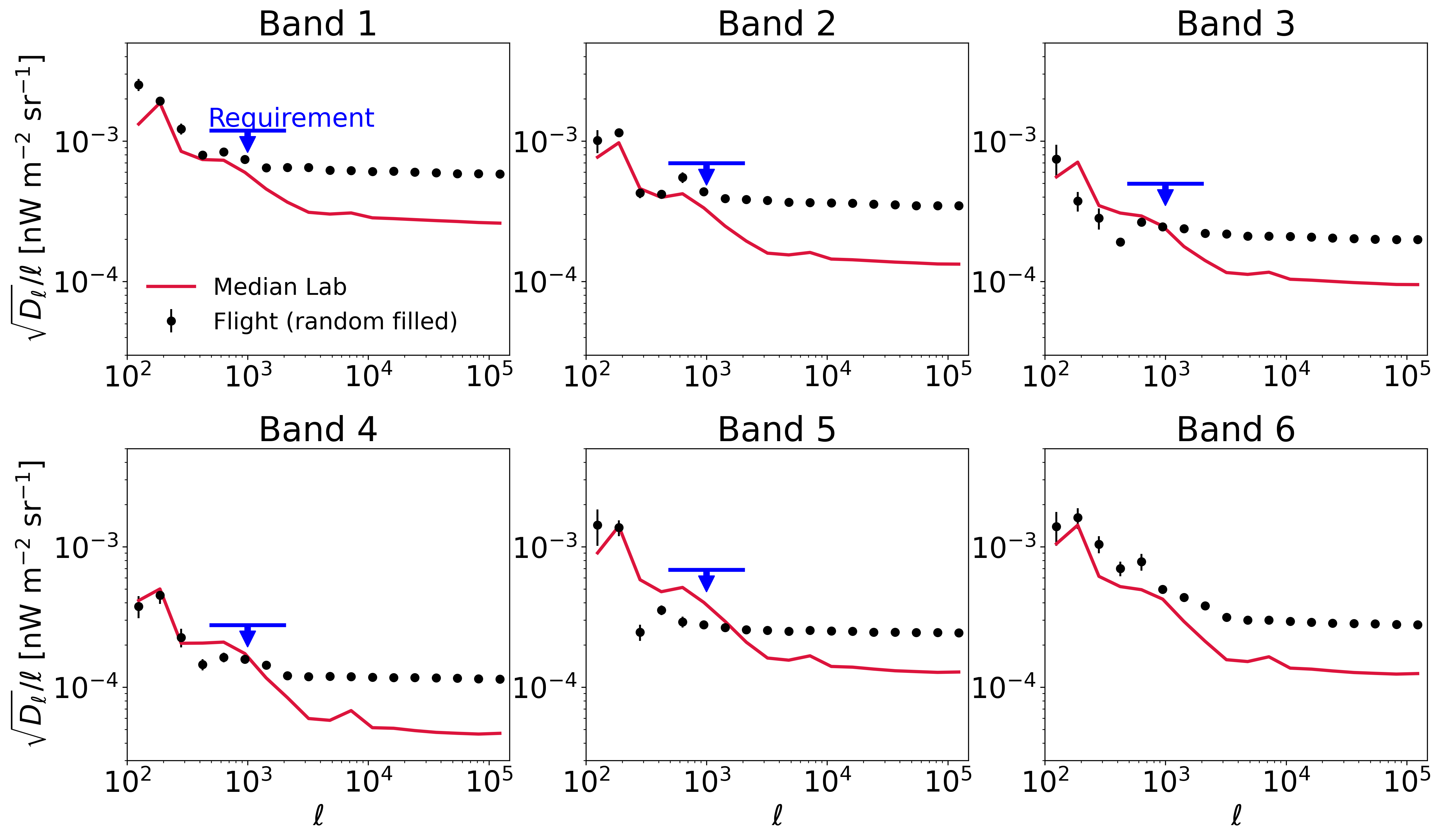}
    \caption{Spatial power spectrum of the difference between image pairs taken during in-orbit checkout (black points).  The red curves give the noise properties measured pre-flight in a dark environment, which show a lower photon noise level but similar behavior at low spatial frequencies.  For this analysis, we subtracted the average signal of the reference pixels from each readout channel.  The required sensitivity for mapping EBL fluctuations, translated to noise in image pairs, is given at multipoles $500 < \ell < 2000$ (blue lines) that are relevant to the angular scale of linear clustering.  Note the galaxy formation noise requirement goes out to 4 $\mu$m and does not include band 6.  The lab measurements showed improved low-frequency performance if we subtract the average signal in the optical pixels of each channel, although this method does remove some astrophysical information.}
    \label{fig:1fnoise}
\end{figure*}

To validate noise stability on orbit, we measured the noise in pairs of consecutive images pointed at the same
location on the sky.  Differencing the two images removes most of the sky signal, though we still must mask
around bright stars that do not perfectly subtract.  The resulting power spectrum of these image differences,
after subtracting the reference pixel averages, meets the statistical sensitivity requirement for the
galaxy formation theme, as shown in Figure \ref{fig:1fnoise}.

\subsection{Sensitivity}
\label{ssec:sensitivity}
\begin{figure*}
\hspace{-1.2cm}
    \includegraphics[trim={0cm 0cm 0cm 0cm}, clip, width=1.15\linewidth]{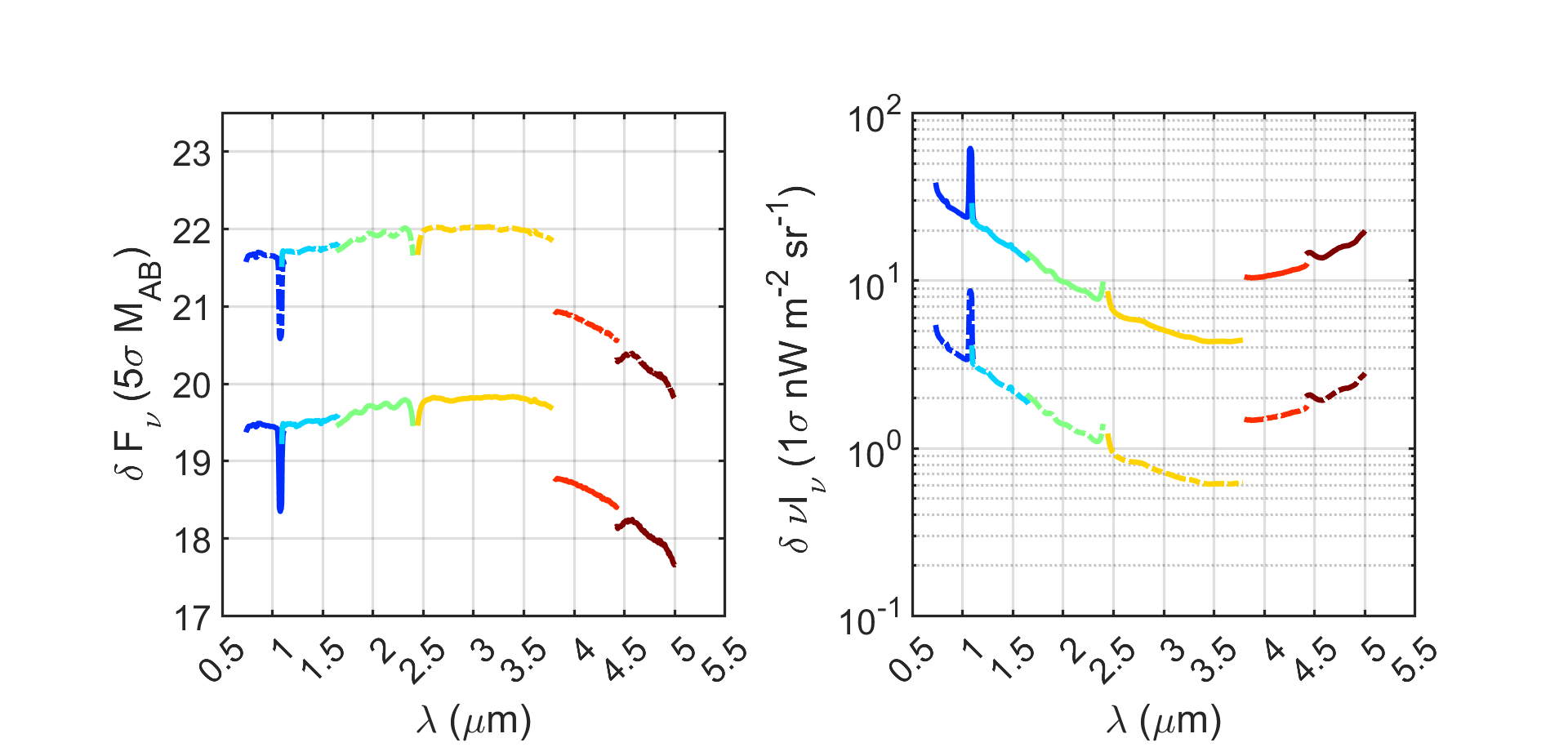}
    \caption{Point source (left) and surface brightness (right) sensitivity given per spectral channel for the all-sky (solid lines) and deep (dashed lines) surveys.  Note that there are 17 spectral channels that provide 17 independent measurements in each band (see Figure \ref{fig:bandc}).  The all-sky sensitivity is based on 4 observations in each spectral channel at the median sky brightness after removing the Galactic plane where $|b| < 25^{\circ}$.  The deep survey assumes 200 observations with the deep field sky brightness at the ecliptic poles and neglects confusion noise.  The point source sensitivity assumes the median $N_{eff}$ in each observation for each spectral channel.  The point source sensitivity is given at $5\sigma$, while the surface brightness is given at $1\sigma$ in each 6\farcs2 SPHEREx pixel.  Note the degraded sensitivity at 1.08 $\mu$m due to photon noise from terrestrial He line emission, and the effect of the beam splitter at 2.4 $\mu$m.  The drop in sensitivity in bands 5 and 6 is due to the higher spectral resolving power in these bands (see Table \ref{table:spectro}).}
    \label{fig:sensitivity}
\end{figure*}

We estimate SPHEREx's point source and surface brightness sensitivity, shown in Figure \ref{fig:sensitivity}. 
We obtain the responsivity from observations of calibration stars, which are connected to all of the pixels
in the focal plane using the estimated flat-field.  We use the per-pixel noise measured under the median sky
brightness to estimate the all-sky sensitivity, and the noise from daily image pairs taken in the deep fields
at low Zodiacal sky brightness to estimate the deep-field sensitivity.  Finally, we use the median $N_{eff}$
in Figure \ref{fig:Neff} at each wavelength to calculate the point source sensitivity.

\begin{figure}
\hspace{-1.55cm}
    \includegraphics[trim={8cm 0.0cm 8cm 0cm}, clip, width=1.4\linewidth]{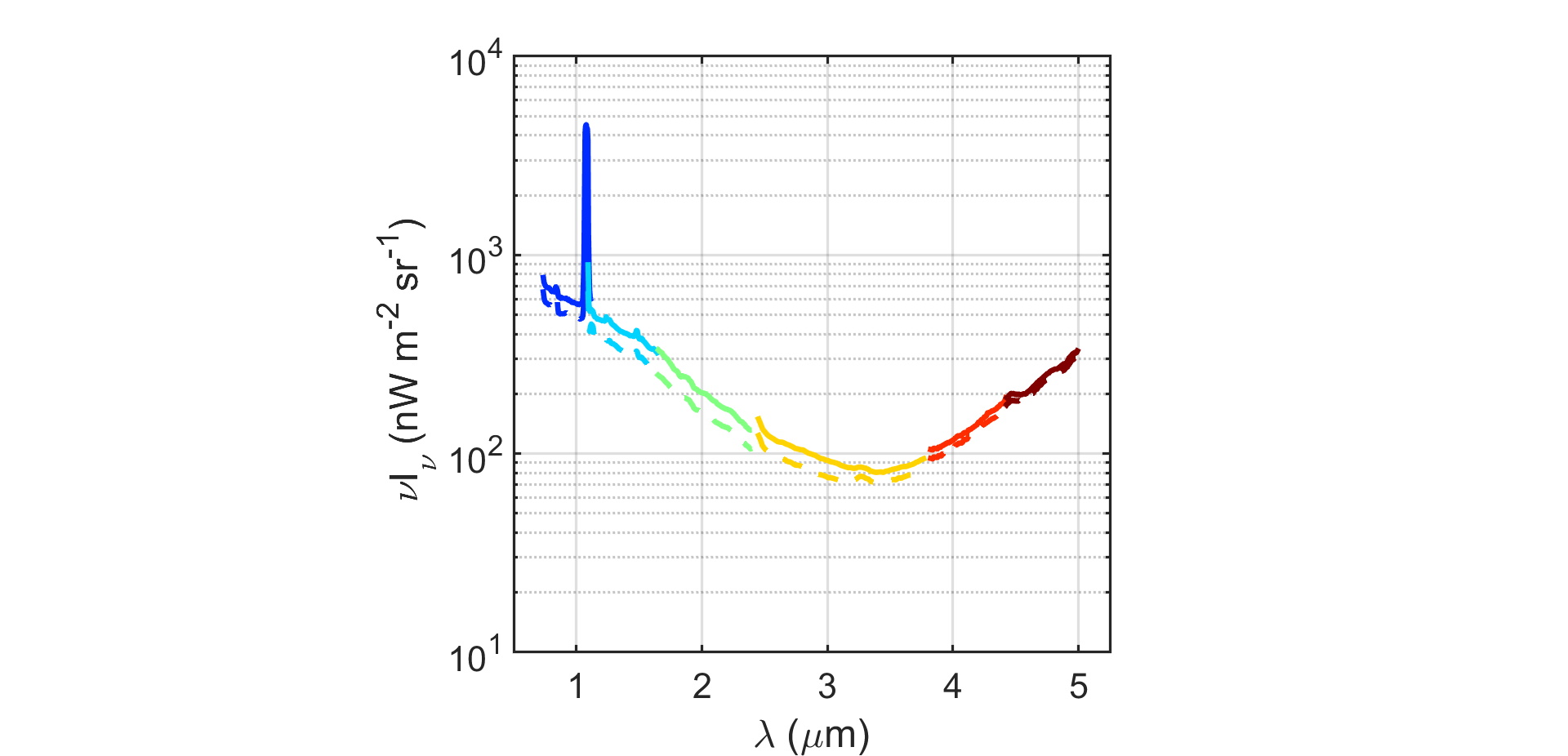}
    \caption{Median sky brightness levels assumed for the all-sky (solid line) and deep-field (dashed line) sensitivities in Figure \ref{fig:sensitivity}.  While Zodiacal light is the dominant component, there are visible contributions from 1.083 $\mu$m He line emission and the interstellar medium.  The slight upturn at 2.4 $\mu$m comes from an imperfect correction for the DBS.  As SPHEREx performance is dominated by photon noise, the sensitivity generally scales as the square root of the sky brightness when considering other regions.}
    \label{fig:mediansky}
\end{figure}

\subsection{Stray Light Response}
\label{ssec:straylight}
We are in the process of quantifying a range of stray light phenomena from in-flight observations \citep{Dowell25}. 
Although there are a few exceptions, the effects and levels are generally in accordance
with a prelaunch design study of the stray light properties.  Within the optical
field of view, we can readily characterize the extended PSF to enable PSF subtraction and source masking.
We observe scattered light from bright stars in narrow regions close to the frame edges, and ghost images
caused by reflections in the beam splitter that are most noticeable near the 2.4 $\mu$m transition.

Outside of the field of view, there is a detectable artifact from stars about $8^{\circ}$ from the field of
view.  The response was initially found prior to the start of science observations by taking images near
the Moon, and was subsequently mapped with bright stars.  While this effect is generally small, we can flag
and remove images where bright stars are near regions of susceptibility as a further mitigation.
We observe features along the left and right edges which appear to be associated with the moon shining
on the exterior of the instrument, although we note the moon does not shine on the instrument during
observations of the deep fields at the ecliptic poles.  Prior to the start of the science survey, we
made a series of observations where the telescope boresight approached the Earth's limb to quantify
the response to earthshine, which was not detected at angles in our current Earth avoidance criteria.

\subsubsection{Voxel Completeness}
Thus far into science operations, SPHEREx is executing a survey plan with $> 99.5$ \% voxel coverage per
survey (see Figure \ref{fig:allsky}) without appreciable data losses, using arrays with an average yield
of 99.15 \% (see Table \ref{table:fpas}).  Transient flags in the detector data averaged 1.1 \% during
the first 3 months of observations, which we expect to gradually drop as solar activity decreases in the solar
cycle.  If we exclude an allocation of affected but unflagged pixels surrounding transient events, we
may lose up to $\sim$4.5 \% of voxels if these cannot be corrected in analysis.  The instrument routinely
observes satellite streaks across the images, however the amount of data loss is relatively small at
$\sim$0.1 \%.  While sources of atmospheric emission (see \S \ref{sssec:atmosphere} - \S \ref{sssec:aurorae})
contaminate some images, these do not appreciably affect data return.

\subsection{Terrestrial Signatures}
\label{ssec:terrestrial}
\begin{figure}
\hspace{-0.2cm}
    \includegraphics[trim={0cm 0 0cm 0cm}, clip, width=0.97\linewidth]{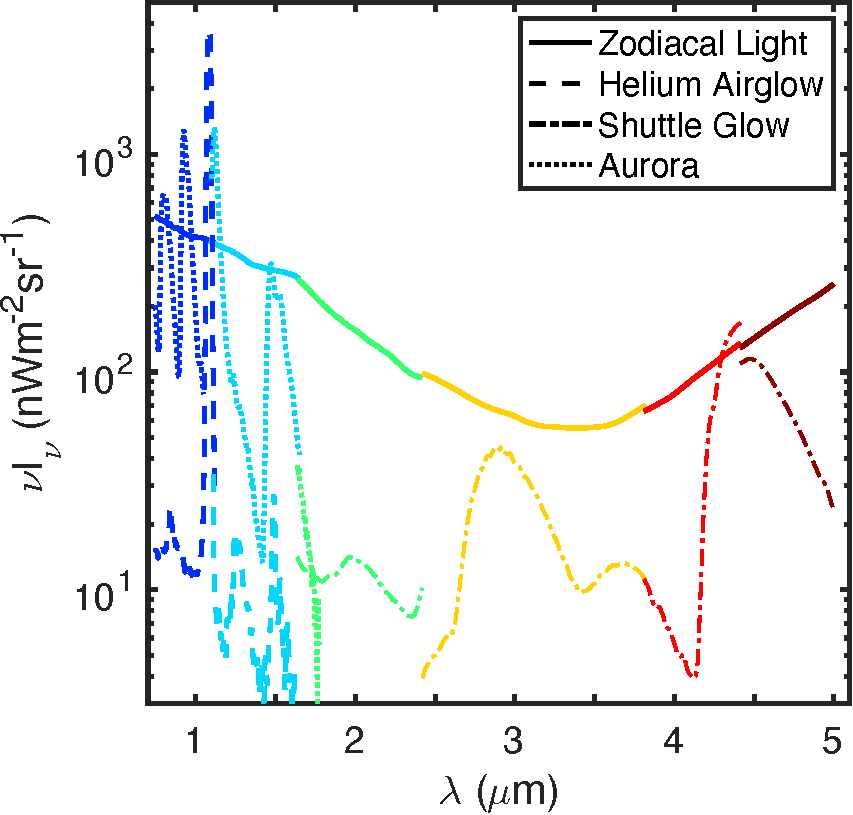}
    \caption{Measured components of the atmosphere including Helium line emission (dashed line), shuttle glow
    (short- and long-dashed line) and aurorae (dotted line).  For comparison, we show the observed spectrum
    of Zodiacal light.  The amplitude of the terrestrial components vary and the levels here are shown when
    the component is prominent.  The colors of the lines indicate the 6 SPHEREx spectral bands.  The small discontinuities in the Zodiacal light between bands are due to uncertainties in the preliminary calibration.  The large step in shuttle glow between bands 5 and 6 appears to be due to its spatial distribution of the emission over the focal plane, being brighter on the long wavelength side.}
    \label{fig:terrsigs}
\end{figure}

SPHEREx observes a number of phenomena associated with its environment in low-earth orbit.  We are working
to identify these phenomena systematically, so they can be mitigated or avoided, depending on the
application.  We have developed spatial templates of the atmospheric features (see Figure \ref{fig:terrsigs})
to aid in regressing residual emission from the spectral images.  We have elected to make images
affected by terrestrial phenomena publicly available, and to keep a running log of affected images that
are identified by data users.

\subsubsection{Atmospheric Emission}
\label{sssec:atmosphere}
We observe bright He line emission \citep{Kulkarni25, Brammer14} at 1.083 $\mu$m in both bands 1 and 2.
The amplitude varies over the
orbit and with time, ranging from a few to 50 times brighter than the Zodiacal light level.  Photon noise
from the line unfortunately reduces sensitivity in this band (see Figure \ref{fig:sensitivity}).
We detect low levels of spectral leakage through the LVFs from the He line, most evident at 0.84, 1.25 and
1.49 $\mu$m.  Other atmospheric lines are much fainter than He but detectable, most notably the atomic
oxygen lines at 0.78, 0.84 and 1.13 $\mu$m.

\subsubsection{Shuttle Glow}
\label{sssec:shuttleglow}
Shuttle glow arises from the spacecraft interacting with the upper atmosphere.  Atomic oxygen striking
the leading surfaces in the direction of travel produces a glow in the optical and infrared 
\citep{Viereck91} due to excited $NO_2$ \citep{Murad98}.  Emission from other molecules have been
proposed in the literature, including $NO, NO^+, OH, CO, CO_2$ and $O_2$, which depend on the composition of
the materials on the spacecraft that chemically react with atomic oxygen.  \cite{Holtzclaw94} measured
the emission spectrum with $\sim0.5~\mu$m resolution of the Z306 black paint used on the SPHEREx telescope,
exposed to a fluence of atomic oxygen.  The authors observe emission peaks at $\sim$2.8 $\mu$m and $\sim$4.4 $\mu$m,
similar to the spectrum in Figure \ref{fig:terrsigs}, which they attribute to vibrational emission from $OH, CO$
and $CO_2$.  As noted in \S \ref{sssec:earthsun}, we observed that shuttle glow increased strongly when the
observatory was tipped forward along the direction of orbital motion.  By restricting the range of zenith
angles in the ram direction, we have reduced shuttle glow so that it is generally fainter than the Zodiacal
sky, though a residual brightness remains.

\subsubsection{Aurorae}
\label{sssec:aurorae}
We occasionally observe auroral line emission from the upper atmosphere associated with solar activity,
especially when SPHEREx is near the polar regions.  The aurorae have prominent features in bands 1 and 2,
with fainter lines visible in bands 3-5.  Although infrequent, bright auroras in their short-wavelength
features can be brighter than the Zodiacal sky.

\subsubsection{Satellites}
SPHEREx frequently observes image streaks produced by Earth-orbiting satellites.  Typically, the streaks are
a few pixels wide, amounting to a small data loss.  We occasionally see a dashed pattern from rotating
satellites.  Bright streaks are flagged on board as transients, and fainter streaks are flagged by the
data pipeline.  We are working to improve the streak identification algorithm to detect streaks near
the noise level, and to better optimize the area flagged along streaks.  We also find that roughly 1\%
of images have bright and diffuse emission that occurs over a fraction of the exposure time, which we
believe are caused by objects in low-earth orbit.

\subsubsection{Energetic Particles}
The H2RG detectors observe a range of phenomena associated with energetic particle interactions.  The
majority of events are single pixels flagged by the on-board SUR algorithm by a jump in the
integrated charge (see \S \ref{sssec:electronics}).  Less frequently, we observe clustered events and
`snowballs' that affect groups of pixels.  We are working to improve the identification of unflagged
pixels near transient events that are also affected by a noticeable increased charge.  The rate of
particle transients varies strongly with solar activity, and often increases when SPHEREx is near the
poles or the SAA.  The mission-averaged rate of flagged transients to date is 1.1 \% of pixels, excluding
images taken in the SAA that are not counted in the survey coverage.  The SUR algorithm returns an estimated
signal from these pixels, although it may be noisy due to the shortened integration.  An additional
$\sim$0.1 \% of pixels are affected in the first 10 s of the integration, but go unflagged by the SUR
routine.

\section{Conclusions}
\label{sec:conclusion}

By mapping the full sky in low-resolution near-infrared spectroscopy, SPHEREx is opening a new capability
in astronomy.  We described mission performance towards meeting requirements on systematic errors
such as gain stability, noise bias, photometric errors, stray light, and dark current, realizing that a full
assessment is beyond the scope of this paper and must be quantified in later scientific publications.
Mission operations have proceeded smoothly to date since the start of science observations on 1 May 2025.
The instrument and spacecraft are performing well, exceeding preflight requirements for
operating temperature, PSF size, pointing stability, point source sensitivity, and surface brightness
sensitivity.

\section{acknowledgements}
\label{sec:acknowledgements}
This publication makes use of data products from the Spectro-Photometer for the History of the
Universe, Epoch of Reionization and Ices Explorer (SPHEREx), which is a joint project of the Jet
Propulsion Laboratory and the California Institute of Technology in the U.S., funded by the National
Aeronautics and Space Administration, and the Korea Astronomy and Space Science Institute (KASI)
in South Korea.  The Korean contribution to SPHEREx was supported by KASI under the Korea AeroSpace 
Administration (KASA).

Part of the research was carried out at the Jet Propulsion Laboratory and the California Institute
of Technology, under a contract with the National Aeronautics
and Space Administration (80NM0018D0004).  Work at Argonne National Laboratory was supported by the U.S.
Department of Energy, Office of High Energy Physics. Argonne, a U.S. Department of Energy Office of Science
Laboratory, is operated by UChicago Argonne LLC under contract no. DE-AC02-06CH11357.

The project thanks the many people who helped develop and implement hardware for the mission, including
but not limited to staff at NASA, JPL, BAE, IPAC, IRSA, KASI, SpaceX, VSFB, Caltech, Viavi, Applied
Aerospace Structures, UC Berkeley SSL, and Teledyne.


\newpage
\bibliography{references}{}
\bibliographystyle{aasjournal}

\section{Appendix}
\label{sec:appendix}

We calculate the minimum pixel size that maintains point source sensitivity with the constraints that the
number of pixels and the observing time are fixed.  The sensitivity to a point source is given by

\begin{equation}
    \Delta F_{\nu} = \sqrt{\frac{N_{eff}}{N_s}}\frac{Rh}{A\eta}\sqrt{\frac{6\nu I_{\nu}A\Omega_{pix}}{5Rh\nu t_i}+\frac{6Q_{CDS}t_s}{t_i^3}},
\end{equation}
where $N_{eff}$ is the effective number of pixels used to measure a point source, $N_s$ is the number of
independent observations, $R =\lambda / \Delta\lambda$ is the resolving power, $h$ is Planck's constant,
$A$ is the mirror area, $\eta$ is the end-to-end optical efficiency, $\nu I_{\nu}$ is the surface brightness
of the sky set by Zodiacal light, $\Omega_{pix}$ is the solid angle of a pixel, $t_i$ is the integration
time, $Q_{CDS}$ is the read noise in a correlated double sample, and $t_{s}$ is the sampling interval in an
SUR integration.  The first term in this equation comes from photon noise, and the second from read noise.  We
note that the read noise term is generally an underestimate, as it only assumes white noise behavior, but
additional terms contribute and become important as the integration time becomes longer.  To be limited by
photon noise, the first term must dominate over the second, ideally with additional margin.

The number of effective pixels in optimal photometry is given by

\begin{equation}
    N_{eff} = \frac{1}{\Sigma p_j^2},
\end{equation}
where $p_{j}$ is the fraction of flux that falls into pixel j.  In the limit where pixels are larger than the
size of the point spread function (PSF), $N_{eff}$ is typically 2-3 on average, depending on the sub-pixel
location of the source for each object.  To maximize sensitivity, one should make the aperture A as large
as possible and the spectral resolving power R as low as possible.  However, the minimum resolving power was
already chosen in \S \ref{ssec:specres} for sufficient redshift accuracy.  Therefore, the remaining free
parameters are aperture and pixel size.

SPHEREx was designed to map the entire sky with a fixed number of detector pixels, which is a non-standard
optimization problem in astronomy.  In this regime, the integration time $t_i$ and the photocurrent $I$ both scale with
the pixel size $\theta_{pix}$ as

\begin{equation}
    I = \frac{\nu I_{\nu}\eta A\Omega_{pix}}{Rh\nu} = I_0 \left(\frac{A}{A_0}\right) \left(\frac{\theta_{pix}}{6.2''}\right)^2,
\end{equation}
and
\begin{equation}
    t_i = t_0 \left(\frac{\theta_{pix}}{6.2''}\right)^2,
\end{equation}
where $I_0$ and $t_0$ are the photocurrent and integration time obtained with the choice of 6\farcs2
pixels, and $A_0$ is the mirror area chosen in our design.  Substituting these into the first equation, and
ignoring small changes in $N_{eff}$ with pixel size, we find

\begin{equation}
    \Delta F_{\nu} = \Delta F_{\nu_0} \left(\frac{A_0}{A}\right)\sqrt{\frac{\Delta I_{\gamma_0}^2\left(\frac{A}{A_0}\right)+\Delta I_{RN_0}^2\left(\frac{6.2''}{\theta_{pix}}\right)^6}{\Delta I_{\gamma_0}^2+\Delta I_{RN_0}^2}},
\end{equation}
where $\Delta I_{\gamma_0}$ and $\Delta I_{RN_0}$ are the uncertainties in the detector current from photon
noise and read noise, respectively, for our default choices of mirror area and 6\farcs2 pixels.
In order to be limited by photon noise, the first term in the numerator must dominate, and for our choice of
parameters in bands 1-4, and the performance of the detectors,
$\Delta I_{\gamma_0} / \Delta I_{RN_0} \sim 3$.  This result means SPHEREx is comfortably limited by photon
noise.  However, given the strong $\left(6.2'' / \theta_{pix}\right)^6$ dependence of the second term,
photon noise will no longer dominate with even modest reductions in pixel size.  One can also see in
the photon noise limit

\begin{equation}
    \Delta F_{\nu} = \Delta F_{\nu_0} \sqrt{\frac{A_0}{A}},
\end{equation}
where increasing the aperture improves sensitivity, but only modestly.

\end{document}